\documentclass[prd,jmp,reprint,onecolumn,superscriptaddress,12pt]{revtex4-2}

\usepackage{bm}
\usepackage{amsfonts}
\usepackage{latexsym}
\usepackage[latin1]{inputenc}
\usepackage{graphicx}
\usepackage{amsmath}
\usepackage[mathscr]{eucal}
\usepackage{palatino}
\usepackage{mathpazo}
\usepackage{textcomp}
\usepackage{float}
\usepackage{booktabs}
\usepackage{dcolumn}
\usepackage{ragged2e}
\usepackage{hyperref}
\hypersetup{colorlinks,citecolor=blue}
\hypersetup{colorlinks=true,linkcolor=red,filecolor=magenta,    urlcolor=blue}
\usepackage{amsmath}
\usepackage{xcolor}
\usepackage[caption=false]{subfig}
\usepackage{commath}
\begin{document}

\title{Scalarization of planar anti-de Sitter charged  black holes in Einstein-Maxwell-scalar theory}

\author{Chatchai Promsiri}
\email{chatchaipromsiri@gmail.com}
\affiliation{Faculty of Science, King Mongkut's University of Technology Thonburi, 126 Prachauthid Road., Bangkok 10140, Thailand}

\author{Takol Tangphati}
\email{takoltang@gmail.com}
\affiliation{Faculty of Science, King Mongkut's University of Technology Thonburi, 126 Prachauthid Road., Bangkok 10140, Thailand}

\author{Ekapong Hirunsirisawat}
\email{ekapong.hir@kmutt.ac.th}
\affiliation{Theoretical and Computational Physics (TCP); Theoretical and Computational Science Center (TaCS), Faculty of Science, King Mongkut's University of Technology Thonburi (KMUTT), Pracha Uthit Road, Bangkok, 10140, Thailand}
\affiliation{Learning Institute, King Mongkut's University of Technology Thonburi (KMUTT), Pracha Uthit Road, Bangkok, 10140, Thailand}

\author{Supakchai Ponglertsakul}
\email{supakchai.p@gmail.com}
\affiliation{Strong Gravity Group, Department of Physics, Faculty of Science, Silpakorn University, Nakhon Pathom 73000, Thailand}

\date{\today}

\begin{abstract}
We construct scalarized planar charged black holes in Einstein-Maxwell-scalar (EMS) theory with the presence of a negative cosmological constant. 
Domains of existence of black hole solutions are given in term of nonminimally coupling constant $\alpha$. Perturbative stability of a  scalarized black hole is investigated by calculating its quasinormal modes. Thermodynamic properties of the scalarized planar solution are also discussed. We observe no evidence of instability of the scalarized black holes. Moreover, we find that scalarized planar charged AdS black holes are thermodynamically preferred over scalar-free solutions in grand canonical and canonical ensembles. The transition between scalar-free solutions and scalarized solutions is found to be the thermal second order phase transition. The transition of these solutions shares some similar features with conductor-superconductor phase transition.

\end{abstract}

\maketitle

\section{Introduction}

It is widely accepted that black holes in general relativity (GR) cannot express other physical quantities rather than mass, electric charge and angular momentum or as John Wheeler simply puts it as a \textit{black hole can have no hair} \cite{gravitation}. The no-hair conjecture states that any stationary black hole solutions of Einstein-Maxwell theory (i.e., electrovacuum) must belong to the Kerr-Newmann family.  For instance, Bekenstein shows noneexistence of asymptotically flat black holes with scalar, vector, and spin-2 meson hair  \cite{PhysRevD.5.1239, *Bekenstein:1972ky, *Bekenstein:1972ny, *PhysRevD.51.R6608}. However, it is found that a black hole can possess macroscopic external degrees of freedom or hairs in several other setups. These include when GR couples with other types of matter fields,  asymptotic structure of spacetime is modified, and gravitational theories beyond Einstein-Maxwell theory, etc. Interestingly, these hairy black holes have much richer physics than their bald counterpart. There are several studies providing a counterexample of the no-hair conjecture, e.g., black holes with Yang-Mills hair and its variants \cite{PhysRevLett.64.2844, *Aichelburg:1992st, *Donets:1992zb, *Kleihaus:1997rb, *Winstanley:1998sn, *PhysRevD.93.064064, *PhysRevD.47.2242, *Ponglertsakul:2016fxj}, hairy black holes in a boxlike boundary \cite{Ponglertsakul:2016wae, *PhysRevLett.116.141101, *PhysRevLett.116.141102, *Sanchis-Gual:2016tcm, *Basu:2016srp, *Peng:2017squ, *Dias:2018yey}, and hairy black holes in modified gravity theories \cite{Brihaye:2015qtu, *BenAchour:2018dap, *Creminelli:2020lxn, *Lee:2021uis, *Erices:2021uyu}. For a nice review on this subject, see \cite{Volkov:2016ehx, Herdeiro:2015waa}.

A bald black hole is able to dynamically develop into a black hole with scalar hair via \textit{spontaneous scalarization}. This usually occurs in the models with nonminimally coupled scalar field. The coupling term of scalar field tends to make the scalar-free black hole solution unstable and leads to the formation 
of a black hole with nontrivial scalar field profile outside its horizon, or a \textit{scalarized black hole (SC BH)}. Spontaneous scalarization is originally considered in scalar tensor theory for neutron star where scalar field is nonminimally coupled to Ricci curvature \cite{PhysRevLett.70.2220} and later extends to rapidly rotating neutron star \cite{PhysRevD.88.084060}. It is found that the scalarized neutron star is energetically favored over the scalar-free solution. Despite this, it is shown in \cite{Hawking:1972qk, PhysRevLett.108.081103} that black holes in scalar tensor theory do not differ from GR, however, scalarized black holes in scalar tensor theory are made possible by surrounding black holes with nonconformally invariant matter \cite{PhysRevD.88.044056, PhysRevLett.111.111101}.

Beyond scalar tensor theory, spontaneous scalarization is also found in extended scalar tensor Gauss-Bonnet (eSTGB) gravity where scalar field is generally coupled to the Gauss-Bonnet curvature term \cite{PhysRevLett.120.131103, *PhysRevLett.120.131104, *PhysRevLett.120.131102, *PhysRevLett.123.011101, *PhysRevLett.123.011101, *PhysRevLett.126.011103, *PhysRevLett.126.011104}. Despite the fact that many studies have been devoted to explore spontaneous scalararization in the eSTGB gravity, however, one of the remaining tasks is to determine the endpoint of instability of the scalar free solutions and dynamical evolution of scalarized solutions. Nonlinear higher curvature terms in the eSTGB gravity render these problems to be challenging. There is, however, a considerably simpler model that also allows for the spontaneous scalarization, i.e., Einstein-Maxwell-scalar (EMS) theory with nonminimally coupling function between scalar field and Maxwell field. A fully nonlinear dynamical evolution from Reissner-Nordstr\"om (RN) black hole into scalarized black hole in the EMS theory is investigated in \cite{PhysRevLett.121.101102}. In this model, scalarized black holes depend on the coupling function between the scalar field and the Maxwell field. In \cite{Fernandes:2019rez}, the dependence of various coupling functions and dynamical features of scalarized black holes are discussed. Moreover, stability and quasinormal modes of scalarized black holes are also investigated by several studies \cite{Myung:2019oua, Myung:2018jvi, Zou:2020zxq, LuisBlazquez-Salcedo:2020rqp}. More interestingly, spontaneous scalarization in the EMS theory in asymptotically de Sitter (dS) \cite{Brihaye:2019gla}, anti de Sitter (AdS)  \cite{Luo:2022roz, *Zhang:2021etr, *Guo:2021zed} and in a cavity \cite{Yao:2021zid} are proposed and studied. 

The seminal papers of Bekenstein \cite{PhysRevD.7.2333} and Hawking \cite{Hawking:1975vcx} suggest that black holes could have entropy and nonzero temperature. Since then, black hole thermodynamics has become one of the most interesting topics to black hole physics communities. 
Thermodynamics of BHs in asymptotically flat spacetime has been often studied, however, there are some unsettled issues about thermal equilibrium configurations.
By putting the BHs in AdS space, Hawking and Page \cite{Hawking:1982dh} find that BHs can be in thermal equilibrium with its surroundings since the AdS boundary acts as a reflecting wall.
The thermodynamics and phase transition of Reissner-Nordstr\"om black hole in AdS (RN-AdS) are studied in \cite{PhysRevD.60.064018, *Chamblin:1999hg, *Burikham:2014gwa}.
Recently, by investigating thermodynamics of asymptotically AdS BHs in the EMS theory in a normal phase space, where the cosmological constant $\Lambda$ is fixed, scalarized solutions are found to be thermodynamically preferred over the RN-AdS BHs in the microcanonical ensemble \cite{Guo:2021zed}.
Interestingly, this system exhibits a reentrant phase transition, which consists of zeroth order and second order types of phase transitions between RN-AdS and SC BHs in some range of parameters.
However, one may consider $\Lambda$ as a thermodynamic variable analogous to pressure $P$ in the first law of BH thermodynamics. 
This framework is called an extended phase space approach \cite{Kastor:2009wy, *Dolan:2012jh, *Kubiznak:2012wp, *Gunasekaran:2012dq, *Kubiznak:2016qmn, *Altamirano:2014tva}.
In this way, phase structure of BHs in the EMS model has been studied in both canonical and grand canonical ensembles \cite{Guo2022Extended}.

It is known that the BHs in asymptotically AdS spacetime can have three different geometries of event horizon with positive, zero and negative curvature constants.
These are called the spherical, planar and hyperbolic BHs, respectively.
The thermodynamics and phase structure of these topological BHs were studied in various theories of gravity, for example see \cite{Birmingham:1998nr, *Emparan:1999gf, *Cai:2001dz, *Cai:2014znn, *PhysRevD.102.024042, *Priyadarshinee:2021rch, *KordZangeneh:2020qeg}.
Intriguingly, thermal properties of AdS BHs and an area law of BH's entropy lead to the development of gauge/gravity duality, which states that the thermodynamics of BHs in a higher dimensional bulk AdS space is holographically dual to a thermal state of gauge theories that living into the AdS boundary \cite{Maldacena:1997re, *Witten:1998zw, *Gubser:1998bc, *Aharony:1999ti}.
By means of gauge/gravity duality, planar BH has received more attention in nongravitational physics communities, such as, condensed matter physics \cite{PhysRevD.78.065034, *Gubser:2005ih, *PhysRevLett.101.031601, *Hartnoll:2008kx, Hartnoll:2009ns, Faulkner:2009wj, Hartnoll:2009sz, Baggioli:2021xuv}, hydrodynamics \cite{Policastro:2001yc, *Son:2007vk, *Bhattacharyya:2007vjd, *Bhattacharyya:2008mz, *Burikham:2016roo, *Baggioli:2020ljz, *Ahn:2022azl, *Baggioli:2022pyb} and quantum information \cite{PhysRevLett.96.181602, *Ryu:2006ef, *Ogawa:2011bz, *Wu:2014xva, *Momeni:2016yts, *Chen:2021lnq}.
In recent years, holographic superconductor in a probe limit with nonminimally coupled EMS theory are considered \cite{Chen:2021fmj, *Mohammadi:2022buy}

In the present paper, we extend the previous studies of \cite{Guo:2021zed, Guo2022Extended} by considering spontaneous scalarization of four dimensional AdS BH with planar horizon in EMS gravity with nonminimal coupling of the scalar and Maxwell fields. 
We establish domain of existence, perturbative stability of scalarized solutions.
The law of BH mechanics demonstrates the mathematical analogy between the dynamics of BH and the law of thermodynamics \cite{Bardeen:1973gs}.
Thus, it is interesting to investigate the spontaneous scalarization in the context of BH thermodynamcis. The differences between spherical and planar horizon of BHs on the scalarization mechanism is also discussed in both mechanical and thermodynamical perspectives. 

This paper is organized as follows. We begin Sec. \ref{section2} with the action for the Einstein-Maxwell-scalar model in AdS spacetime and present its field equations along with the boundary conditions for the numerical method. Moreover, the bifurcation line that presents the emergence of scalarized planar AdS charged black hole is studied. In Sec. \ref{section3}, the numerical results of the scalarized planar AdS charged black hole are demonstrated with the analysis of the stability of the scalarized solutions. In Sec. \ref{section4}, the study of the thermodynamics properties of the scalarized planar AdS charged black hole is presented in the grand canonical ensemble (fixed potential) and canonical (fixed charge) ensemble. We also present the Euclidean regularized action technique for eliminating the divergences at the boundary and from the electromagnetic (EM) field. Finally, we summarize the results and discuss the novel findings in the conclusions.

\section{Einstein-Maxwell-Scalar model in AdS spacetime}\label{section2}

In this section, we discuss basics equations involving the EMS-AdS theory. We will derive equations of motion and boundary conditions that allow an existence of scalarized solutions. Then we explore the origin of instability that leads to scalarization.

\subsection{The model}\label{subsec:model}

We consider the model such that massless scalar field ($\varphi$) is minimally coupled to gravity but nonminimally coupled to gauge field $A_{\mu}$. The action is given by
\begin{align}
    S &= \frac{1}{16\pi} \int d^4 x \sqrt{-g} \bigg[R - 2\Lambda -2 \nabla_{\mu}\varphi\nabla^{\mu}\varphi - \mathcal{G}(\varphi)F_{\rho\sigma}F^{\rho\sigma} \bigg], \label{action}
\end{align}
where $F_{\rho\sigma}=\nabla_{\rho}A_{\sigma} - \nabla_{\sigma}A_{\rho}$ and coupling between the scalar field and the gauge field is denoted by $\mathcal{G}(\varphi)$. The cosmological constant relates to the AdS radius $L$ by $\Lambda = -3/L^2$. 

Varying \eqref{action} with respect to $g^{\mu\nu},A^{\mu}$ and $\varphi$, we obtain the following equations of motion
\begin{align}
    R_{\mu\nu} - \frac{g_{\mu\nu}R}{2} + \Lambda g_{\mu\nu} &= 2T_{\mu\nu}, \label{EFE}\\
    \nabla_{\mu}\left(\mathcal{G}F^{\mu\nu}\right) &= 0, \label{MW}\\
    \nabla^{\mu}\nabla_{\mu}\varphi &= \frac{1}{4}\frac{d\mathcal{G}}{d\varphi}F_{\rho\sigma}F^{\rho\sigma}, \label{KG} 
\end{align}
where the energy-momentum tensor is defined by
\begin{align}
   T_{\mu\nu} &= \partial_{\mu}\varphi\partial_{\nu}\varphi - \frac{1}{2}g_{\mu\nu}\partial^{\rho}\varphi\partial_{\rho}\varphi + \mathcal{G}\left(g^{\rho\sigma}F_{\mu\rho}F_{\nu\sigma} - \frac{1}{4}g_{\mu\nu}F_{\rho\sigma}F^{\rho\sigma}\right).
\end{align}
Remark that, the nonminimally coupling $\mathcal{G}$ cannot take an arbitrary form. Instead, we must choose this coupling function that allows an existence of the scalar free solution. This means when $\varphi=0$, the action \eqref{action} admits Reissner-Nordstr\"om-AdS solution. This puts a condition on function $\mathcal{G}$ i.e., $\mathcal{G}(0)=1$ and $\frac{d\mathcal{G}(0)}{d\varphi}=0$. In the absence of the gauge field, the action \eqref{action} admits black hole solutions with scalar hair \cite{PhysRevD.64.044007, *Winstanley:2002jt, *Sudarsky:2002mk}.

\subsection{Equations of motion}

Here we consider spacetime metric with planar symmetry. The spacetime metric, the gauge field and the scalar field take the following forms
\begin{align}
    ds^2 &= -N(r)e^{-2\delta(r)}dt^2 + N(r)^{-1}dr^2 + \frac{r^2}{L^2}\left(dx^2+dy^2\right), \label{metric} \\
    A_{\mu} &= \{V(r),0,0,0\}, \\
    \varphi &= \phi(r).
\end{align}

We define mass function $m(r)$ as 
\begin{align}
    N(r) &\equiv -\frac{2m(r)}{r} + \frac{r^2}{L^2}.
\end{align}
Putting these into equations of motion \eqref{EFE}--\eqref{KG}, we obtain 
\begin{align}
    m' &= \left(\frac{r^3}{2L^2} - m\right)r\phi'^2 + \frac{Q^2}{2r^2\mathcal{G}}, \label{EFE1}\\
    \delta' &= - r\phi'^2, \label{EFE2} \\
     V' &= -\frac{Q \text{e}^{- \delta}}{r^2 \mathcal{G}}, \label{MW1} \\
     0 &= \left(\frac{r^4}{L^2} - 2 r m\right)\phi'' + \left(\frac{4r^3}{L^2} - 2 m - \frac{Q^2}{r\mathcal{G}}\right) \phi' + \frac{Q^2}{2r^2\mathcal{G}^2}\frac{d\mathcal{G}}{d\phi} \label{KG1}.
\end{align}

We denote derivative with respect to radial coordinate $r$ by $\prime$ e.g., $m'=\frac{dm}{dr}$. $Q$ is constant of integration which can be associated to black hole's charge (see Appendix\ref{Komar integral}). When $\phi=0$, the RN-AdS with planar symmetry is the solution to these equations. To obtain a scalarized planar AdS charged black hole, we must solve these nonlinear differential equations \eqref{EFE1} with appropriated boundary conditions.

\subsection{Boundary conditions}

We assume that there exists regular event horizon located at $r=r_+$ i.e., $N(r_+)=0$. The field functions $\{m,\delta,V,\phi\}$ are expected to be finite near the event horizon and at spatial infinity. Therefore, we expand the field functions near the event horizon accordingly
\begin{align}
    m &= m_0 + m_1 (r-r_+) + ..., \\
    \delta &= \delta_0 + \delta_1 (r-r_+) + ..., \\
    V &= V_0 + V_1 (r-r_+) + ..., \\
    \phi &= \phi_0 + \phi_1 (r-r_+) + ...
\end{align}
where equation of motions \eqref{EFE1}--\eqref{KG1} allow us to determine the following
\begin{align}
    m_0 &= \frac{r_+^3}{2L^2}, \quad
    m_1 = \frac{Q^2}{2r_+^2\mathcal{G}(\phi_0)}, \\
    \delta_1 &= -r_+\phi_1^2, \\
    V_1 &= -\frac{e^{-\delta_0}Q}{r_+^2\mathcal{G}(\phi_0)}, \\
    \phi_1 &= \frac{Q^2L^2 }{2r_+\mathcal{G}(\phi_0)\left(Q^2L^2 - 3 r_+^4 \mathcal{G}(\phi_0)\right)}\frac{d\mathcal{G}(\phi_0)}{d\phi}.
\end{align}
Here, we use the gauge freedom to set $V_0=0$. At spatial infinity, we obtain
\begin{align}
    m &= M - \frac{Q^2}{2r} + ..., \label{Bcinf1}\\
    \phi &= \frac{\phi_f}{r^3} + ..., \label{Bcinf2}\\
    \delta &= \frac{3\phi_f^2}{2r^6} + ..., \label{Bcinf3}\\
    V &= \Phi + \frac{Q}{r} + ..., \label{Bcinf4}
\end{align}
Here, the parameter $M$ is related to the Komar mass $E$ of black hole (see Appendix \ref{Komar integral}). The leading order of scalar field at spatial infinity is denoted by  constant $\phi_f$. 
The electrostatic potential $\Phi$ is defined as the difference between the gauge field at infinity and event horizon
\begin{eqnarray}
\Phi = A_t(\infty)-A_t(r_+) = V(\infty)-V(r_+).
\end{eqnarray}
Since our equations of motion \eqref{EFE1}--\eqref{KG1} has shifted symmetry in $V$, therefore we can add arbitrary constant such that $V(r_+)=V_0=0$. 
Thus the electric potential of black hole solution becomes
\begin{eqnarray}
\Phi = A_t(\infty)=V(\infty).
\end{eqnarray}
Remark that one obtain an explicit form of $\Phi$ by integrating \eqref{MW1}
\begin{eqnarray}
    \Phi &=& -\int_{r_+}^{\infty} dr \frac{Q e^{-\delta}}{\mathcal{G}(\phi)r^2}. \label{eqPhi}
\end{eqnarray}
This equation serves as a good check on our numerical results. 

\subsection{Tachyonic instability}

Before discussing on scalarized solution, it is useful to understand what drives bald planar AdS charged black hole away from its stability. Thus in this subsection, we consider a linear scalar perturbation on fixed background of planar AdS charged black holes. We obtain such a solution by choosing 
\begin{align}
    \phi(r) = 0, \quad \delta(r)=0, \quad N(r) = -\frac{2M}{r} + \frac{r^2}{L^2} + \frac{Q^2}{r^2}, \quad A_t = \frac{Q}{r}.
\end{align}
Equation governing scalar perturbation on curved background can be obtained by linearization of \eqref{KG} with a small perturbation $\delta \varphi$ 
\begin{align}
    \left(\square - \mu_{\text{eff}}^2 \right) \delta \varphi &= 0, \label{linearizedKG}
\end{align}
where
\begin{align}
    \mu_{\text{eff}}^2 = -\frac{d^2\mathcal{G}(0)}{d\phi^2}\frac{Q^2}{2r^4}.
\end{align}
It turns out that the second derivative of the nonminimally coupling function plays a role as the effective mass of scalar field. In asymptotically flat spacetime, tachyonic instability arises if $\mu_{eff}^2 < 0$. For asymptotically AdS spacetime, if $\mu_{eff}^2 < \mu_{BF}^2$ where $\mu_{BF}^2 = -\frac{9}{4L^3}$ is the Breitenlohner-Freedman bound \cite{BREITENLOHNER1982249}, the tachyonic instatilbity occurs. We therefore choose the nonminimally coupling such that  $\frac{d^2\mathcal{G}(0)}{d\phi^2} > 0$. Throughout this work, we particularly consider the coupling in the following form
\begin{align}
    \mathcal{G} &= e^{\alpha \phi^2},
\end{align}
where $\alpha$ is a positive constant. Thus the effective mass squared is $\mu_{\text{eff}}^2 = -\frac{\alpha Q^2}{r^4}$. This form of coupling function also satisfies the requirements discussed in the Sec. \ref{subsec:model}. For other forms of the coupling functions, for instance, hyperbolic, power law, and fraction are investigated thoroughly in \cite{Fernandes:2019rez}.

We expand the small perturbation $\delta \varphi$ as
\begin{align}
    \delta \varphi &= U(r)e^{i\left(k_1 x + k_2 y\right)},
\end{align}
where $k_1,k_2$ can be considered as wave vector in planar direction. Therefore Eq. \eqref{linearizedKG} can be expressed in the form
\begin{align}
    \left(r^2NU'\right)' - \left[\frac{\alpha Q^2}{r^2} - L^2 \Vec{k}^2\right]U = 0,
\end{align}
where $\Vec{k}^2\equiv k_1^2+k_2^2$. The above equation is solved by assuming that the radial function $U$ is regular at the event horizon and smoothly vanishing at spatial infinity. For a given, $Q=0.4,L= 4, \Vec{k}=0$, we solve this equation for each $\alpha$. As a results, we obtain a \textit{bifurcation line} which marks the location where scalarized solutions bifurcate from planar AdS charged black hole. See Fig.\ref{fig:solspace} below for an example plot. We remark that in Einstein-scalar-Gauss-Bonnet-AdS theory, a scalar perturbation on scalar-free spherical BH is comparatively easier leading to tachyonic instability than those with planar horizon \cite{Guo:2020zqm, *Kiorpelidi:2022kuo}. 

\section{Scalarized planar AdS charged solutions}\label{section3}

To obtain scalarized solution, we solve \eqref{EFE1}--\eqref{KG1} with boundary conditions as discussed in the previous section. We apply numerical shooting method where $\phi_0$ and $\delta_0$ are chosen such that the boundary conditions at spatial infinity are satisfied. For a given value of the event horizon $r_+$, black hole charge $Q$, AdS radius $L$ and coupling constant $\alpha$, 
we numerically integrate \eqref{EFE1}--\eqref{KG1} from $(r_+ + \epsilon)$ to some certain distance ($r_{\infty}$) where $\epsilon$ is set to $10^{-9}$. We search for $\phi_0$ and $\delta_0$ that make $\{m,\delta,V,\phi\}$ behave asymptotically as \eqref{Bcinf1}--\eqref{Bcinf4}. We obtain $M$ and $\Phi$ by identifying $M=m(r_{\infty}), \Phi=V(r_{\infty})$. We find that the numerical value of $\Phi$ agrees very well with \eqref{eqPhi}.

\begin{figure}[H]
    \includegraphics[width = 5.4cm]{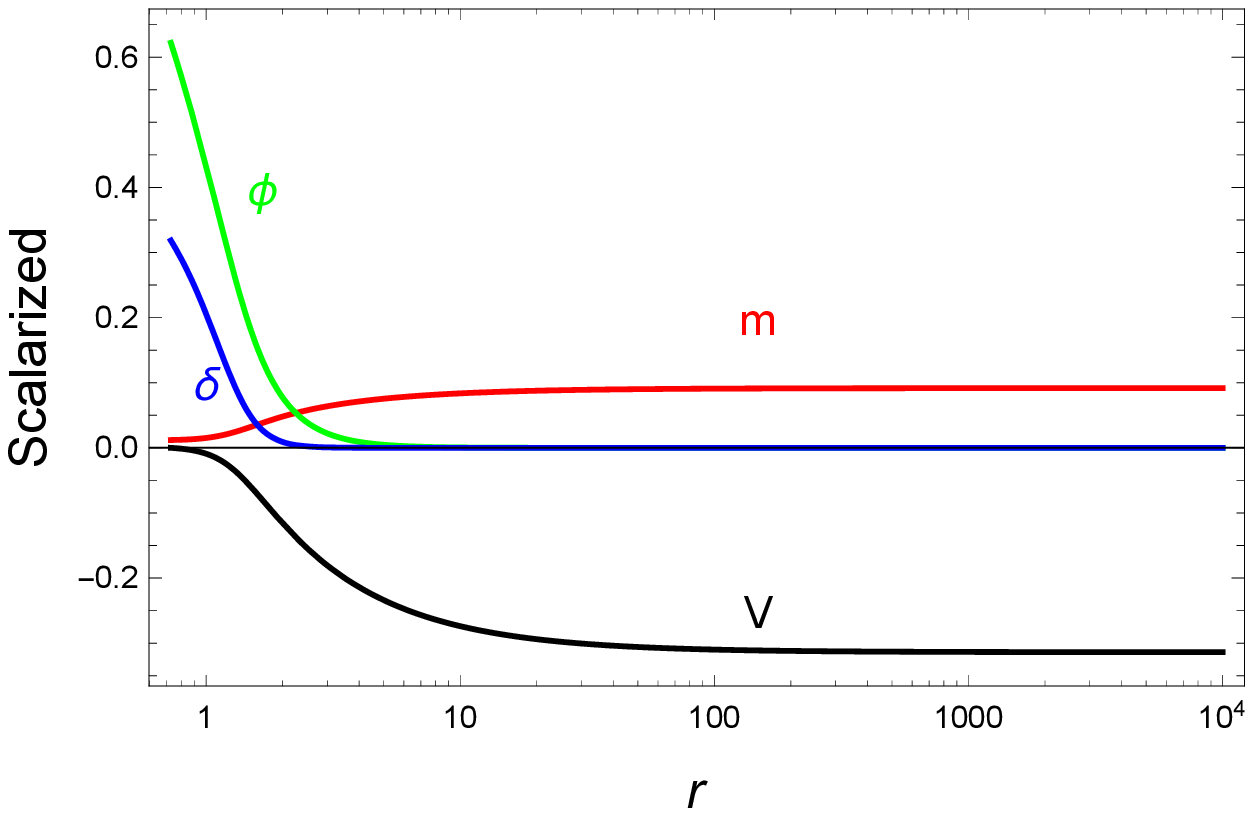}
    \includegraphics[width = 5.4cm]{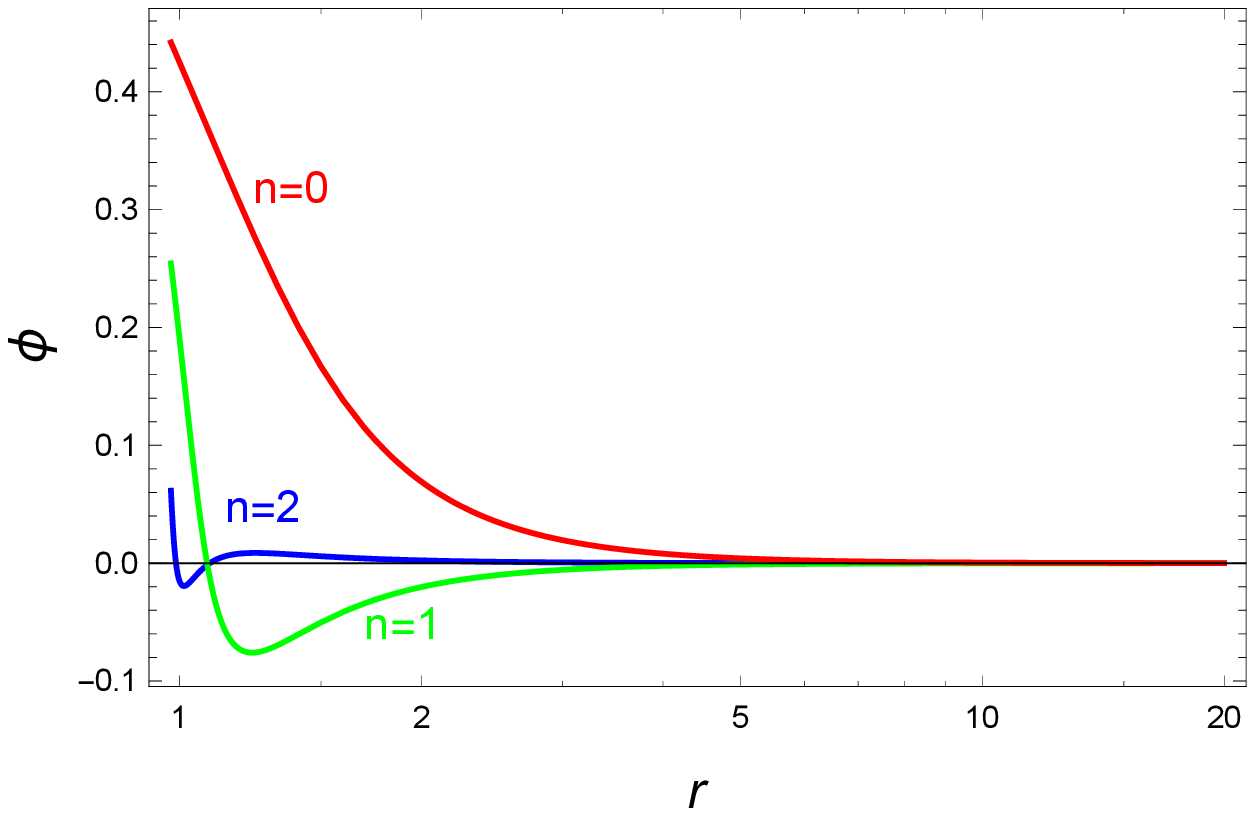}
    \includegraphics[width = 5.4cm]{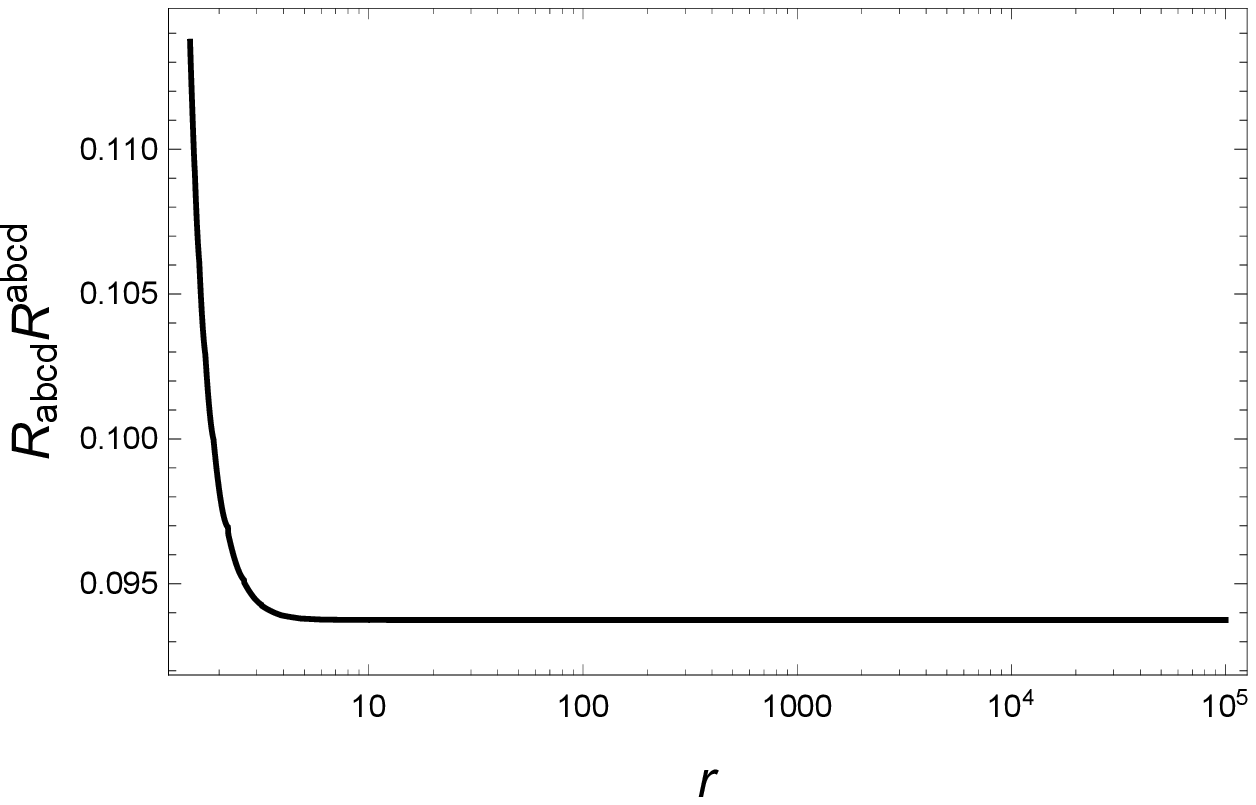}
    \caption{ Left: Example plots of the field functions for $\alpha$=9.3 and $M$=0.092, $q$=4.35. Middle: The profiles of scalar field for $n$=0, $n$=1, $n$=2 modes with $\alpha$=10 and $q$=3.60, 3.62 and 3.91 respectively. Right: Plot the Kretschmann scalar as a function of r for scalarized black hole geometry with $\alpha$=18.7 and $M$=0.152, $q$=2.63.}
    \label{fig:sols}
\end{figure}

To illustrate its behaviors, the plots of scalarized planar AdS charged black holes as the functions of the radius $r$ are displayed with setting $Q=0.4,~L=4$. In Fig~\ref{fig:sols}, we display example plots of scalarized planar AdS charged black holes. 
In the left panel, the free parameters are found to be $\phi_0=0.622,~\delta_0=0.318$. We find that the black hole mass and electrostatic potential are $0.092$ and $-0.314$ respectively. It is obvious that the scalar field profile develops in the exterior region of the black hole horizon and decreases rapidly as $r$ is larger . The scalar field profiles can be characterized by number of node $n$ in the exterior region of black hole. These are shown in the middle panel. For the rest of this work, we shall 
particularly focus on the $n=0$ mode. Lastly, the right panel indicates that there is no essential singularly anywhere outside the black hole's horizon since the Kretschmann scalar $R_{abcd}R^{abcd}$ is always finite for $r>r_+$.

\begin{figure}[ht]
    \centering
    \includegraphics[width = 11cm]{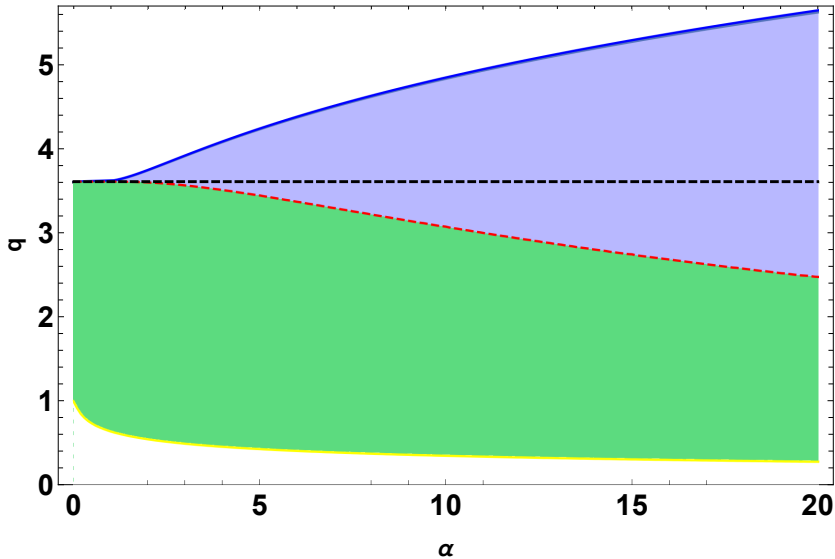}
    \caption{Domain of existence of scalarized planar AdS charged black hole. The red dashed line is the bifurcation line and the dashed black line is the black hole extremal line. The shaded area above the bifurcation line is the domain of existence of scalarized solutions. The upper limit of this domain is critical line (solid blue line). The colored area below the bifurcation line is the area where planar AdS charged black hole exists. All the shaded area is displayed only when $\mu_{eff}^2<\mu_{BF}^2$.
    }
    \label{fig:solspace}
\end{figure}

Solution space of scalarized solution is illustrated in Fig~\ref{fig:solspace}. The solution space is displayed in $\alpha-q$ plane where $q\equiv\frac{Q}{M}$ is defined as the reduced charge.
In this plot, we only express the region where tachyonic instability occurs i.e., $\mu_{eff}^2<\mu_{BF}^2$. 
Here we only evaluate $\mu_{eff}^2$ at event horizon since it takes the smallest possible value. The domain of existence of scalarized solution is bounded by the bifurcation line (red dashed line) and critical line (solid blue line). 
The bifurcation line indicates the location in phase space where scalarized solutions bifurcates from bald black holes. 
The maximum value of $q$ for each $\alpha$ form the critical line where scalarized black hole ceases to exist. 
We find no solution above the critical line. 
Remark that, the scalarized black holes and planar AdS black holes coexist in the region bounded by the bifurcation line and the extremal line. 
On the extremal line the black holes have zero Hawking temperature. In this plot, the extremal line is located at $q=3.604$ or $M=0.11098$. This also implies that above the extremal line the mass of scalarized solutions are lower than those of planar AdS charged black holes. 
From this plot the area of existence increases with coupling constant $\alpha$. We remark that the domain of existence plot of planar black hole is qualitatively similar to those of spherically symmetric solution \cite{Guo:2021zed}. 
The difference is that in spherical symmetric setup the extremal line locates at $q=1$. 

\begin{figure}[ht]
    \centering
    \includegraphics[width = 11 cm]{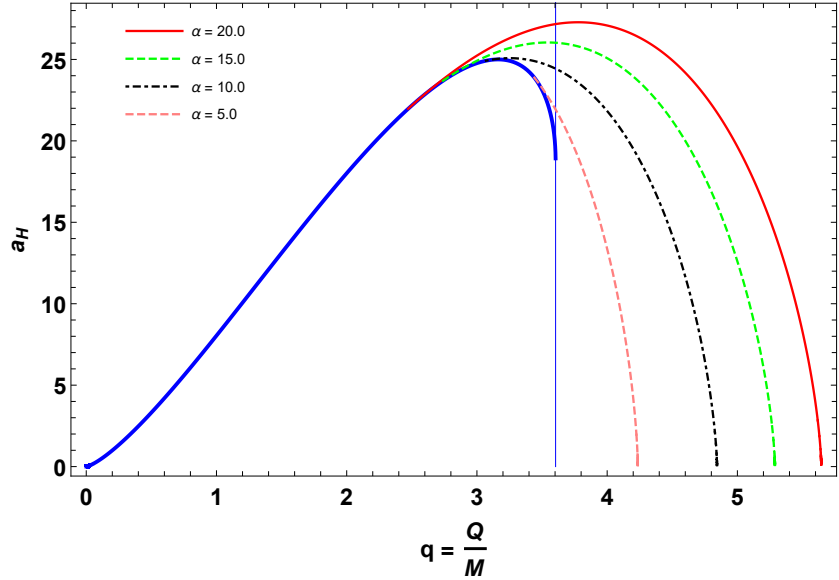}
    \caption{
    The reduced horizon areas $a_H$ against the reduced charge $q$ for scalarized planar AdS BHs in the EMS gravity is plotted with $\alpha = 5, 10, 15$ and $20$ corresponding to dashed pink, dash-dotted black, dashed green and solid red, respectively, compared with the case of the bald AdS charged BH (solid blue). Vertical (solid blue) line represents the extremal point in this case.}
    \label{fig:aH_q}
\end{figure}

In thermodynamics, it is known that the most preferred state is the state with the largest entropy.
Thus, if two states have different entropy, the state with higher entropy is preferred over the other.
According to Bekenstein \cite{PhysRevD.7.2333}, entropy of BHs is proportional to its surface area at horizon radius, i.e., $S_\text{BH}\sim A_H$, where $S_\text{BH}$ is the Bekenstein-Hawking entropy and
the event horizon area of planar BH from the metric tensor is given by $A_H=\mathcal{V}_2r^2_+/L^2$.
Here $\mathcal{V}_2$ denotes the spatial extension in the $xy$ plane. 
In this case, we introduce the reduced horizon area $a_H$, which measures the event horizon area with respect to the area at $r=2M$, i.e., $a_H=r^2_+/4M^2$.
Thus we display the reduced horizon area as a function of $q$ in Fig~\ref{fig:aH_q}. At small $q$, only planar AdS charged black holes exists. Beyond the extremal point $q= 3.604 $ (marked by the vertical line), the planar AdS charged black holes do not exist. When $q$ reaches certain number, scalarized planar black hole bifurcates from the bald one. The bifurcation points for $\alpha=5,~10,~15, \text{ and } 20$ are at $q=3.438,~3.065,~2.735, \text{ and } 2.466$, respectively. Remarkably, we observe that there exists a region where planar AdS black holes co-exist with the scalarized solutions. It is clear that in the co-exist region scalarized solutions are entropically favored over planar AdS charged black holes since their reduced horizon areas are relatively larger than those of planar AdS charged black holes. This agrees with spherical symmetric situation where scalarized black holes are found to be globally stable in microcanonical ensemble \cite{Guo:2021zed}.


\subsection{Stability of scalarized solutions}
In this subsection, we explore linear stability of scalarized black holes. We consider time dependent linear perturbations in EMS theory. Therefore, we expand the field functions $\{N,\delta,V,\phi\}$ 
\begin{align}
    N(t,r) &= N_0(r) + \epsilon_1 N_1(r)e^{-i\Bar{\omega}t}, \\
    \delta(t,r) &= \delta_0(r) + \epsilon_1 \delta_1(r)e^{-i\Bar{\omega}t}, \\
    V(t,r) &= V_0(r) + \epsilon_1 V_1(r)e^{-i\Bar{\omega}t}, \\
    \varphi(t,r) &= \phi_0(r) + \epsilon_1 \phi_1(r)e^{-i\Bar{\omega}t}, 
\end{align}
with frequency $\Bar{\omega}$ and $\epsilon_1$ is order parameter. Here $\delta_0$ and $\phi_0$ are not free parameters but rather represent metric function and scalar function at equilibrium. Thus at first order (in $\epsilon_1$), \eqref{EFE}--\eqref{MW} imply the following,
\begin{align}
    N_1 &= -2rN_0\phi_0' \phi_1, \\
    \delta_1' &= -2r\phi_0' \phi_1', \\
    V_1' &= \frac{Qe^{-\delta_0}}{r^2\mathcal{G}(\phi_0)^2}\left(\mathcal{G}(\phi_0)\delta_1 +  \frac{d\mathcal{G}(\phi_0)}{d\phi_0}\phi_1\right).
\end{align}
We redefine $\phi_1 \equiv \psi/r$ and define tortoise coordinate $dr_{\ast} = e^{\delta_0}N_0^{-1} dr$. Therefore, the perturbed Klein-Gordon equation can be expressed in the Schr\"odinger-like form
\begin{align}
    -\frac{d^2\psi}{dr_{\ast}^2} + \left(V_{\Bar{\omega}} -\Bar{\omega}^2 \right)\psi &=0, \label{ReggeWheeler}
\end{align}
where
\begin{align}
    V_{\Bar{\omega}} &= N_0e^{-2\delta_0}\left[\frac{3-6r^2\phi_0'^2}{L^2} - \frac{Q^2\left(1 - 2 r^2\phi_0'^2\right)}{r^4\mathcal{G}} - \frac{N_0}{r^2} - \frac{Q^2}{2r^4\mathcal{G}^2}\left(4r\phi_0'\frac{d\mathcal{G}}{d\phi_0}+\frac{d^2\mathcal{G}}{d\phi_0^2}\right)\right].
\end{align}

\begin{figure}[ht]
    \centering
    \includegraphics[width = 13 cm]{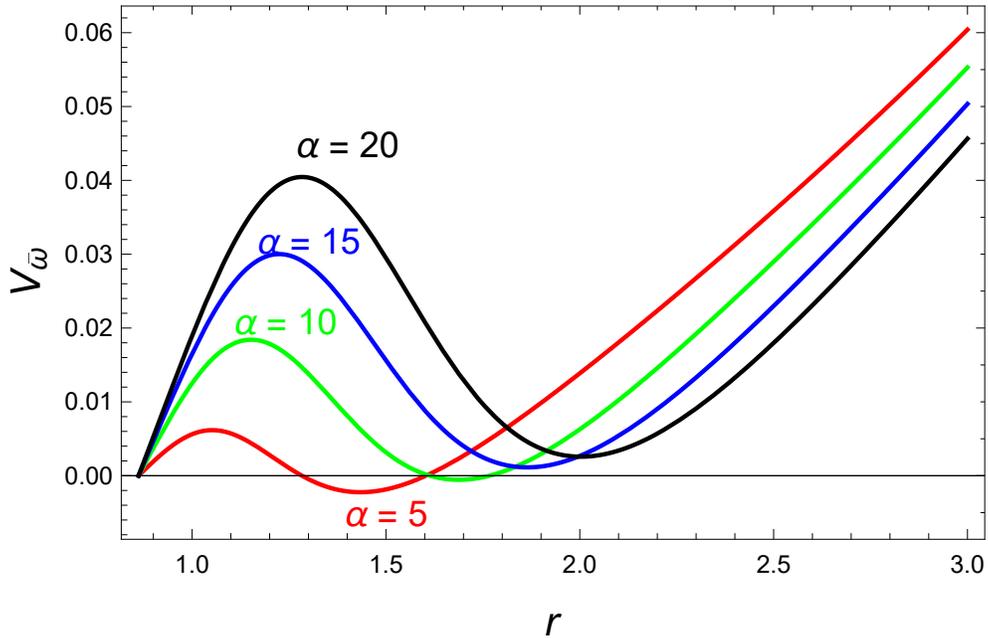}
    \caption{The plot of perturbation potential against radial coordinate $r$ with $Q = 0.4$ and $L = 4$ for various value of $\alpha$. The event horizon locates at $r_+=0.862$.}
    \label{fig:Veff}
\end{figure}

Clearly, the perturbation potential $V_{\Bar{\omega}}$ vanishes at event horizon and positively diverge as we approach spatial infinity. This is illustrated in Fig~\ref{fig:Veff} where the potential is plotted for various value of $\alpha$. Equation \eqref{ReggeWheeler} will have no bound state if $V_{\Bar{\omega}}$ is positive everywhere. The existence of bound state leads to unstable mode with $\Bar{\omega}^2<0$. From Fig~\ref{fig:Veff}, there exists a region where $V_{\Bar{\omega}}<0$ at small $\alpha$. However, as we increase $\alpha$, the potential exhibits no negative region anywhere outside the event horizon. Nevertheless, the existence of negative potential region does not sufficiently leads to development of instability \cite{Fernandes:2019rez,Zou:2019bpt}. Stability of scalarized solution can be analyzed using the S-deformation technique \cite{Kimura:2017uor, *Kimura:2018whv}.

\begin{figure}[ht]
    \centering
    \includegraphics[width = 8 cm]{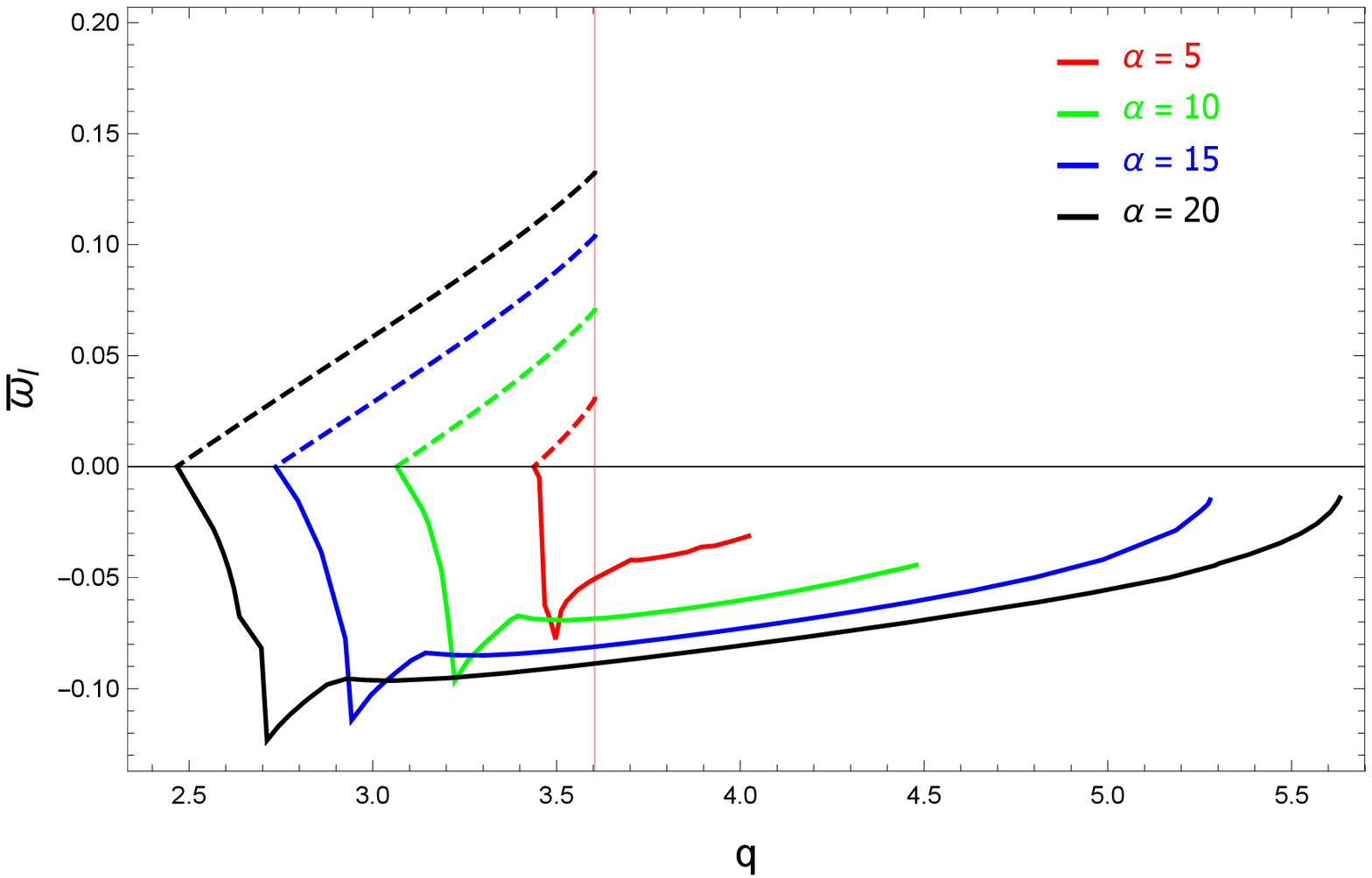}
    \includegraphics[width = 8 cm]{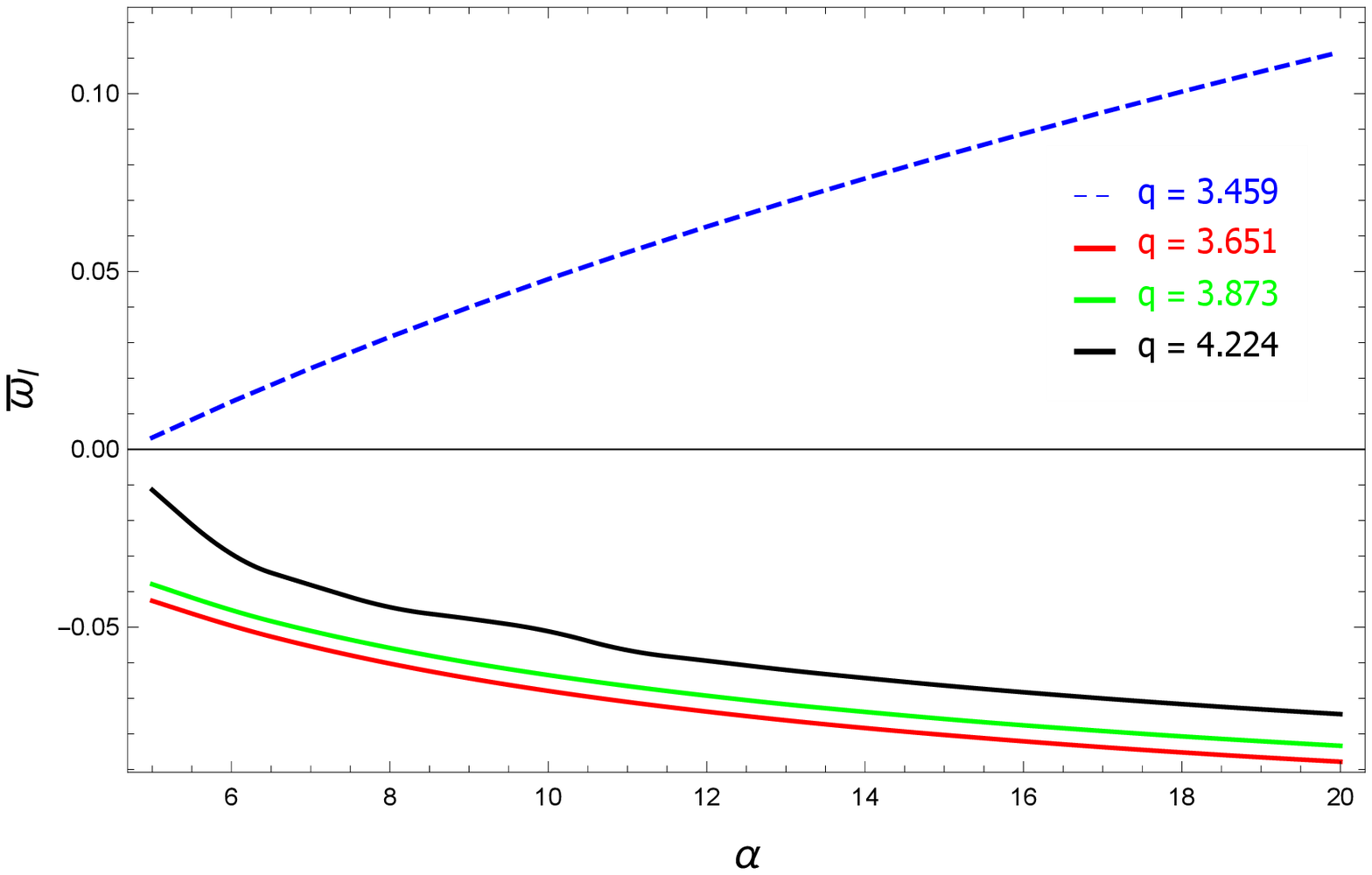}
    \caption{  Imaginary part of lowest-lying quasinormal frequency ($\bar{\omega}_I$) plots against charge $q$ (Left) and coupling constant $\alpha$ (Right). The colored solid lines represent $\bar{\omega}_I$ of scalarized black holes where dashed lines are planar AdS charged black holes. The red vertical line in the left figure is charge at extremal limit $q=3.604$. }
    \label{fig:qnm}
\end{figure}

Furthermore, stability of scalarized solution can be determined by calculating quasinormal modes (QNMs) \cite{Konoplya:2011qq,Kokkotas:1999bd}. We investigate whether the solutions are stable against the linear scalar perturbation \eqref{ReggeWheeler}. For asymptotically AdS spacetime, QNMs are defined to be the modes that satisfy purely ingoing wave at the horizon and vanishing at infinity i.e.,
\begin{align}
    \psi(r\to r_h) \sim e^{-i\bar{\omega}r_{\ast}},~~~~~\psi(r\to\infty)\sim 0. \label{bcs}
\end{align}
The frequency $\bar{\omega}$ corresponding to these modes are quasinormal frequencies. As a consequence of the boundary conditions, the frequencies are complex and discrete. Imaginary part of $\bar{\omega}$ will reveal whether the perturbations are stable (exponentially decay) or unstable (exponentially growth). Thus we numerically solve scalar perturbation \eqref{ReggeWheeler} equation with background numerical nodeless solutions (as mentioned earlier in this section) and imposing boundary conditions \eqref{bcs}. The aim is to obtain the quasinormal frequencies as a function of charge $q$ and nonminimally coupling constant $\alpha$. This is done by using built-in function $NDSolve$ in Wolfram's \textit{Mathematica}. In addition, we implement psuedospectral method in order to obtain the quasinormal frequencies. We refer the readers to \cite{Jansen:2017oag} for a nice introduction and \textit{Mathematica}'s package of this method.

In Fig.~\ref{fig:qnm}, we illustrate an imaginary part of quasinormal frequencies $\bar{\omega}_I$ as a function of $q$ and $\alpha$. In these plots, $\bar{\omega}_I$ of scalarized black holes are displayed by the solid curves where those of RN-AdS are shown by the dashed curves. For each fixed value of $\alpha$, we plot scalarized solutions from the smallest possible $q$ i.e., $q$ at the bifurcation line to the largest possible value ($q$ at the critical line). We observe that the quasinormal frequencies are always negative. Negativity of $\Bar{\omega}_I$ implies that these scalarized solutions are linearly stable. The larger the $\alpha$, the more stable these solutions are. More interestingly, $\Bar{\omega}_I$ develops nontrivial behavior as a function of $q$. At small $q$, it appears that the solutions become more stable as $q$ is increased. However, beyond a certain value of $q$, the solutions are less stable as $q$ is increased. We note that similar trends are also observed in \cite{Guo:2021zed, Gan:2019ibg} for spherically symmetric solutions. Despite this, we find no evidence of unstable modes of scalarized black holes. Moreover, QNMs of RN-AdS are also computed. We find that as $q$ increases, the RN-AdS becomes more unstable as expected. The RN-AdS becomes most unstable when we reach extremal limit denoted by red vertical line. In addition, the right figure of Fig.~\ref{fig:qnm}, we plot $\bar{\omega}_I$ as a function of $\alpha$ for various values of $q$. We notice that when $q$ is increases the $\bar{\omega}_I$ becomes more positive. While increasing $\alpha$, the scalarized solution become more stable. Similar to the previous case, we find that the RN-AdS solutions are unstable against scalar perturbation.

\section{Thermodynamics}\label{section4}

In this section, we investigate the thermodynamics and phase structure of scalarized black holes with the planar horizon in the grand canonical ensemble and canonical ensemble \cite{Guo:2021zed,Guo2022Extended}. 


\subsection{Action calculation}

To analyze the thermodynamics behaviors of black holes in the grand canonical ensemble with fixed electric potential $\Phi$ or canonical ensemble with fixed electric charge $\Tilde{Q} = \frac{\mathcal{V}_2 Q}{4\pi L^2} $ (see Appendix A). We study the Euclidean action of the solution in an imaginary time $t\rightarrow i\tau$ by identifying the period $\beta_\text{H}=1/T_\text{H}$ with the Hawking temperature $T_\text{H}$ \cite{PhysRevD.15.2752}. In the semiclassical approximation, the thermal partition function is given by 


\begin{eqnarray}
    \mathcal{Z} \sim \text{e}^{-S^{E}_\text{on-shell}},
\end{eqnarray}
where $S^{E}_\text{on-shell}$ defined as the on-shell Euclidean action.  Nevertheless, the action \eqref{action} diverges as spacetime volume increases. Thus we must renormalize the action to eliminate the divergences from the asymptotic AdS spacetime.

We introduce the Euclidean regularized action $S^E_R$ including the Euclidean bulk action $S^E_\text{bulk}$ from \eqref{action}, Gibbons-Hawking boundary term $S^E_{\text{GH}}$ for eliminating the divergence at the boundary, the counterterm $S^E_{\text{ct}}$ for eliminating the divergence from the AdS asymptotic at the boundary, and $S^E_{\text{surf}}$ for eliminating the divergence of the EM field,
\begin{eqnarray}
    S^E_R = S^E_\text{bulk} + S^E_{\text{GH}} + S^E_{\text{ct}} + S^E_{\text{surf}}. \label{Total Euclidean action}
\end{eqnarray}
Note that in the above equation, we have changed the time coordinate $t$ into the Euclidean time $\tau = i t$. For convenience, in the Euclidean spacetime, we will denote $d^4 x = d\tau dr dx dy$. The Euclidean bulk and the boundary terms are given by
\begin{eqnarray}
    S^E_\text{bulk} &=& -\frac{1}{16\pi} \int d^4x \sqrt{g} \bigg[R + \frac{6}{L^2} -2 \nabla_{\mu}\varphi\nabla^{\mu}\varphi - \mathcal{G}(\varphi)F^2 \bigg], \label{action_Euclidean} \\
    S^E_{\text{GH}} &=& -\frac{1}{8 \pi} \int d^3 x \sqrt{\gamma^{(3)}} \Theta, \\
    S^E_{\text{ct}} &=& \frac{1}{8 \pi} \int d^3 x \sqrt{\gamma^{(3)}} \left( \frac{2}{L} + \frac{L}{2} \mathcal{R} \right), \label{Sct} \\
    S^E_{\text{surf}} &=&  -\frac{1}{4 \pi} \int d^3 x \sqrt{\gamma^{(3)}} \mathcal{G}(\varphi) F^{\mu \nu} n_{\mu} A_{\nu}, 
\end{eqnarray}
where $\gamma^{(3)}$ is the determinant of the induced metric on the hypersurface at $r \rightarrow \infty$, $\Theta$ is the trace of the extrinsic curvature ${\Theta^{\mu}}_{\mu}$, $\mathcal{R}$ is the scalar curvature of the induced metric $\gamma_{ij}^{(3)}$, and $n_{\mu}$ is the unit normal vector on the hypersurface. 

Now we consider the Euclidean bulk term $S^E_\text{bulk}$ along with the Einstein field equation in \eqref{EFE} and its trace. The Euclidean bulk term can be written as 
\begin{eqnarray}
    S^E_\text{bulk} &=&  \frac{1}{16 \pi} \int d^4 x \sqrt{g} \left[ \frac{6}{L^2} + \mathcal{G}(\varphi) F^2 \right]. \label{Eucidean bulk}
\end{eqnarray}
With the $\tau\tau$ component of the field equation in Eq.~\eqref{EFE}, we obtain the identity as follows,
\begin{eqnarray}
    R_{\tau \tau} + \frac{3}{L^2} N e^{-2\delta} &=& -\mathcal{G}(\varphi) N V'^2. \label{Rtt}
\end{eqnarray}
The field strength tensor square is given by
\begin{eqnarray}
F^2 = -2e^{2\delta} V'^2 \label{Field strength}.
\end{eqnarray}
Plugging Eqs. \eqref{Rtt} and \eqref{Field strength} into \eqref{Eucidean bulk}, the Euclidean bulk term becomes
\begin{eqnarray}
    S^{E}_\text{bulk} &=& - \frac{1}{16 \pi} \int d^4x \sqrt{g} \left[4 e^{2\delta} \mathcal{G}(\varphi) V'^2 + \frac{2 e^{2\delta}}{N} R_{\tau \tau} \right], \nonumber \\
    &=& - \frac{1}{16 \pi} \int^{\beta_H}_0 d\tau \int dx dy \int^{\infty}_{r_+} dr \sqrt{e^{-2\delta} \frac{r^4}{L^4}} \left[ 4 e^{2\delta} \mathcal{G}(\varphi) V'^2 + \frac{2 e^{2\delta}}{N} R_{\tau \tau} \right], \nonumber \\
    &=& - \frac{\mathcal{V}_2}{4 \pi T_H} \int^{\infty}_{r_+} dr \frac{r^2}{L^2} \left[ e^{\delta} \mathcal{G}(\varphi) V'^2 + \frac{1}{2N} e^{\delta} R_{\tau \tau}  \right]. \nonumber
\end{eqnarray}
We substitute the $\tau \tau$ component of the Ricci tensor, which directly calculated from the metric tensor in Eq. \eqref{metric}, into the above equation. One can integrate out the coordinate $r$ to obtain
\begin{eqnarray}
 S^{E}_\text{bulk} &=& \frac{1}{T_\text{H}} \left[ \frac{\mathcal{V}_2}{16\pi L^2}r^2e^{-\delta}\left( N' - 2 N \delta' \right)\big|_{r\rightarrow \infty} - T_\text{H} S_\text{BH} - \tilde{Q} \Phi \right].
    \label{1st_Euclidean_final}
\end{eqnarray}
Here, the Hawking temperature of the scalarized black hole is
\begin{eqnarray}
T_\text{H} = \frac{1}{4 \pi} N'(r_+) e^{-\delta(r_+)}.
\end{eqnarray}
The Bekenstein-Hawking entropy is given in the form of the area law as
\begin{eqnarray}
S_\text{BH} = \frac{A_\mathcal{H}}{4} = \frac{\mathcal{V}_2r_+^2}{4L^2}.
\end{eqnarray}
Note that the horizon area can be calculated from
\begin{eqnarray}
A_\mathcal{H} = \int d^2x\sqrt{\gamma^{(2)}} = \frac{\mathcal{V}_2 r^2_+}{L^2},
\end{eqnarray}
where $\gamma^{(2)}$ is a determinant of the induced metric associated to the horizon surface and $\mathcal{V}_2 = \int dx dy$ is the spatial extension of black hole in $\mathbb{R}^2$.

The first term in \eqref{1st_Euclidean_final} diverges at $r \rightarrow \infty$. To remove this divergence, we make use of the Gibbons-Hawking term to cancel them out. The extrinsic curvature is defined by
\begin{eqnarray}
    \Theta_{\mu\nu} \equiv \frac{1}{2} {h^{\alpha}}_{\mu} {h^{\beta}}_{\nu} \left( \nabla_{\alpha} \sigma_{\beta} + \nabla_{\beta} \sigma_{\alpha} \ \right).
\end{eqnarray}
The metric on the hypersurface $h_{\mu \nu}$ is defined as follows,
\begin{align}
    h_{\mu \nu} \equiv g_{\mu \nu} - \sigma_{\mu} \sigma_{\nu},
\end{align}
where $\sigma_{\mu} = (0, 1/N^{1/2}, 0, 0)$ is the unit normal vector pointing outward from the hypersurface. The trace of the extrinsic curvature can be expressed explicitly
\begin{eqnarray}
    \Theta = \frac{N'}{2\sqrt{N}} - \delta' \sqrt{N} + \frac{2\sqrt{N}}{r}.
\end{eqnarray}
Now the Gibbons-Hawking boundary term is,
\begin{eqnarray}
    S^E_{\text{GH}} &=& -\frac{1}{8 \pi} \int d^3 x \sqrt{\gamma^{(3)}} \Theta, \nonumber \\
    &=& -\frac{1}{T_H} \left[\frac{\mathcal{V}_2}{16\pi L^2}r^2e^{-\delta}\left( N' - 2 N \delta' \right) + \frac{\mathcal{V}_2}{4\pi L^2} e^{-\delta} \left( - 2 M + \frac{r^3}{L^2} \right) \right]_{r\rightarrow \infty}.
\end{eqnarray}
The first term in the above equation cancels out the divergence in the bulk action \eqref{1st_Euclidean_final}.
The second term in the above equation leads to another divergence. However, one finds that the counterterm $S^E_{\text{ct}}$ yields
\begin{eqnarray}
    S^E_{\text{ct}} &=& \frac{\mathcal{V}_2}{4 \pi L^2 T_\text{H}} e^{-\delta} \left( \frac{r^3}{L^2} - M \right) \Bigg|_{r \rightarrow \infty},
\end{eqnarray}
where $\mathcal{R}$ vanishes with planar symmetry. The surface boundary action $S_\text{surf}^E$ vanishes when the potential $\Phi$ is fixed at the boundary since the field strength of the gauge field is zero on this surface. Therefore the grand potential is given by
\begin{eqnarray}
\Omega = T_H S^E_R = E - T_HS_\text{BH} - \tilde{Q}\Phi. \label{Gibbs}
\end{eqnarray}
On the other hand, if we fix the electric charge $Q$ of the black hole instead, the surface term now becomes nonvanishing, i.e.,
\begin{eqnarray}
    S^E_{\text{surf}} &=&  -\frac{1}{4 \pi} \int d^3 x \sqrt{\gamma^{(3)}} \mathcal{G}(\varphi) F^{\mu \nu} n_{\mu} A_{\nu}, \nonumber \\
    &=& \frac{\mathcal{V}_2}{4 \pi T_\text{H} L^2}Q \Phi, \nonumber \\ 
    &=& \frac{1}{T_\text{H}}\tilde{Q}\Phi . \label{surface}
\end{eqnarray}
Therefore, it contributes to the on-shell action in Eq.\eqref{Total Euclidean action}. In this case, the free energy is the Helmholtz free energy which can be written in the following form
\begin{eqnarray}
    F = T_\text{H}S_\text{R}^E = E - T_\text{H}S_\text{BH}. \label{Helmholtz}
\end{eqnarray}

\subsection{Fixed potential}
\begin{figure}[H]
    \centering
    \includegraphics[width = 8. cm]{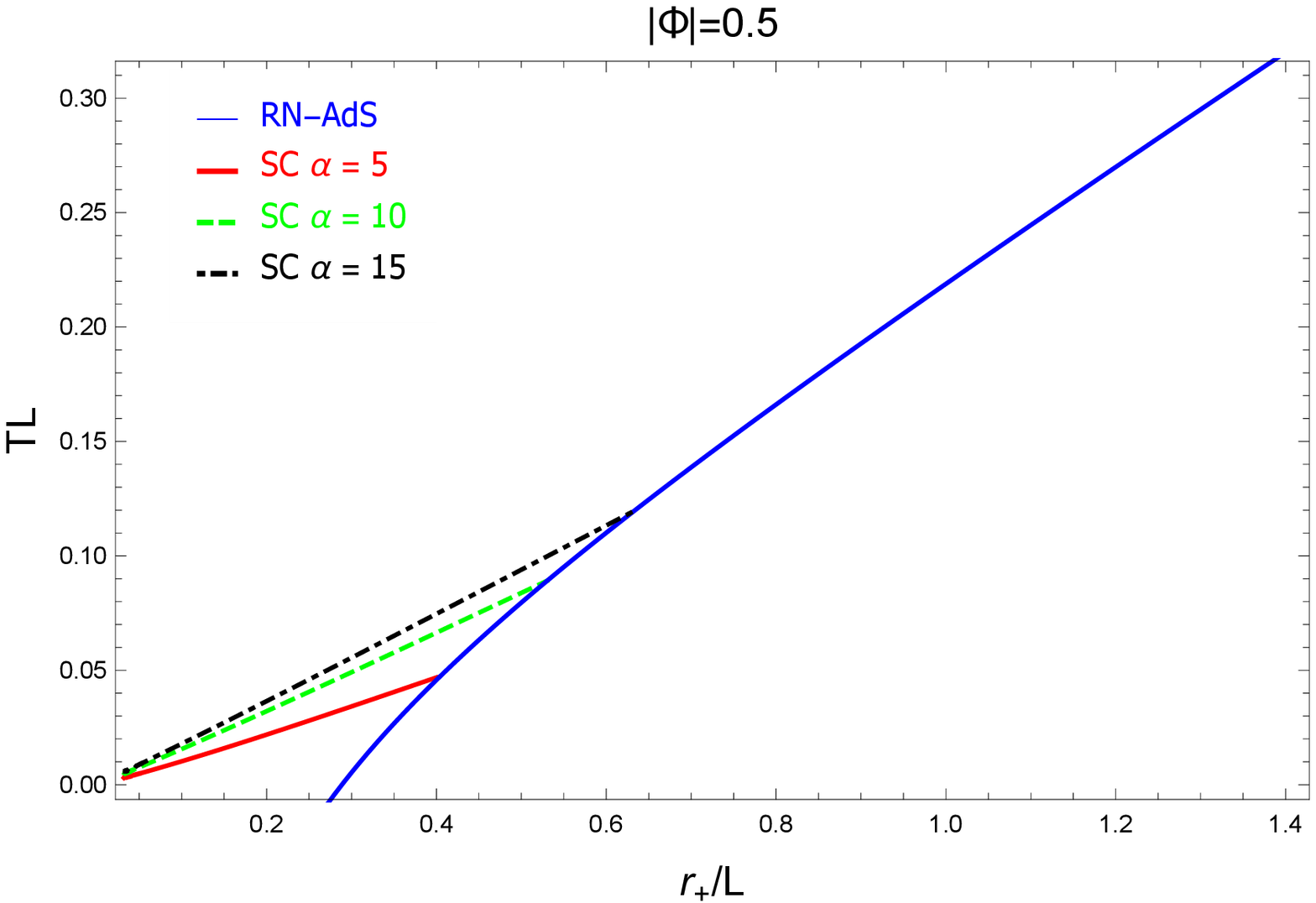}
    \includegraphics[width = 8. cm]{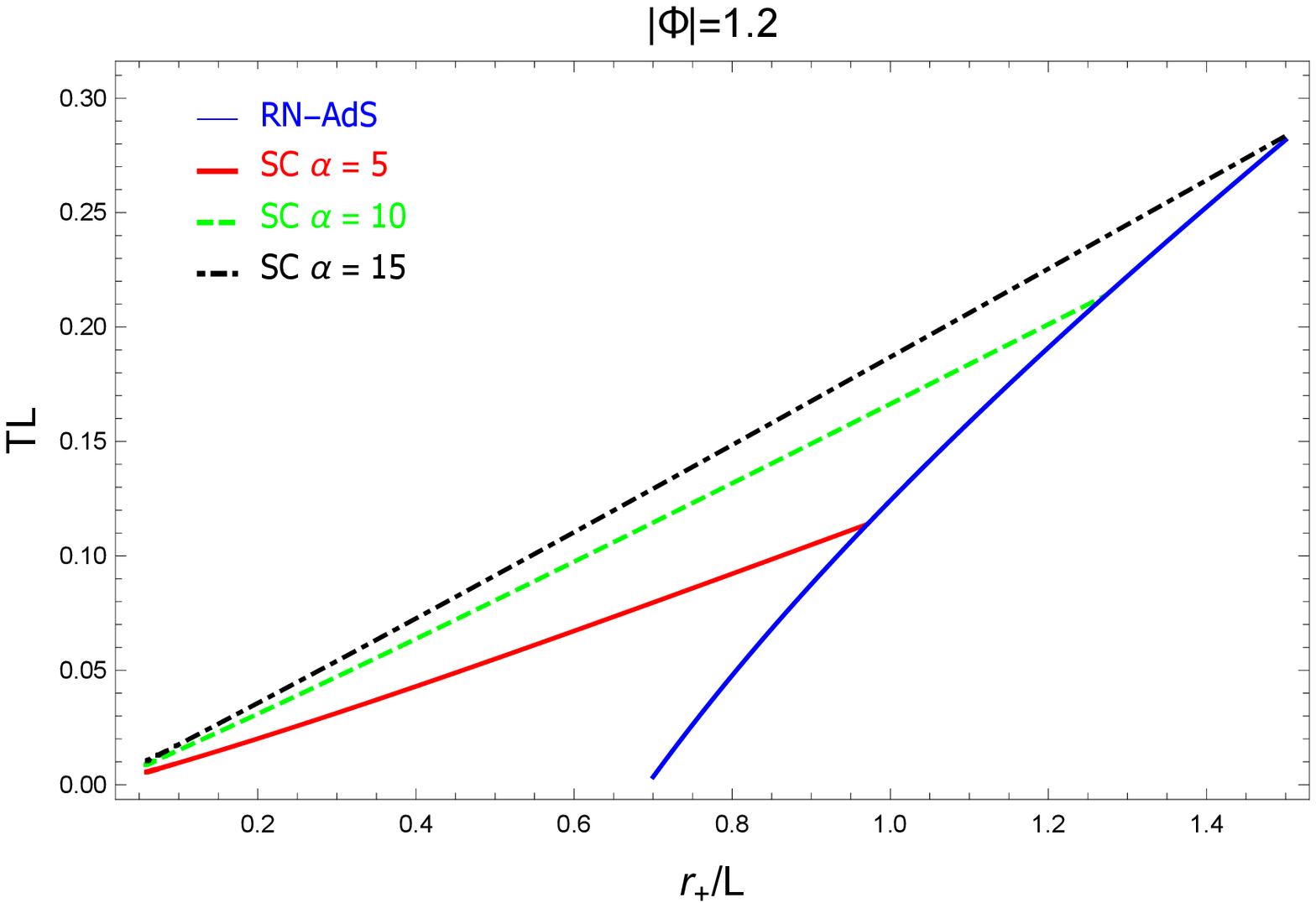}
    \caption{Reduced temperature $TL$ plots against reduced horizon radius $r_+/L$ with $L=1$. Left: for fixed $|\Phi|=0.5$ Right: for fixed $|\Phi|=1.2$. The solid blue and red lines are RN-AdS and scalarized solution with $\alpha=5$. The dashed green and dot-dashed black lines are scalarized solution with $\alpha=10$ and $15$ respectively. }
    \label{fig: Temp vs r+}
\end{figure}

 The behaviors of the reduced temperature $TL$ with reduced horizon radius $r_+/L$ are illustrated in Fig.~\ref{fig: Temp vs r+}. In these plots, we fix the electric potential $|\Phi|=0.5$ and $1.2$ in the left and right panels, respectively. At low temperature, there are two branches of black holes with planar horizon: RN-AdS BH (solid blue curve) and SC BH (solid red, dashed green and dot-dashed black curves which correspond to coupling constant $\alpha =5, 10$ and $15$ respectively). In both BH's configurations, the Hawking temperature is proportional to BH's horizon radius, however, the SC BH has higher temperature than the RN-AdS BH until its temperature increases to $T_B$, where $T_B$ represents the temperature at the bifurcation point. For $T>T_B$, SC BH disappears and there is only one BH branch which is RN-AdS.
   
\begin{figure}[H]
    \centering
     \includegraphics[width = 8. cm]{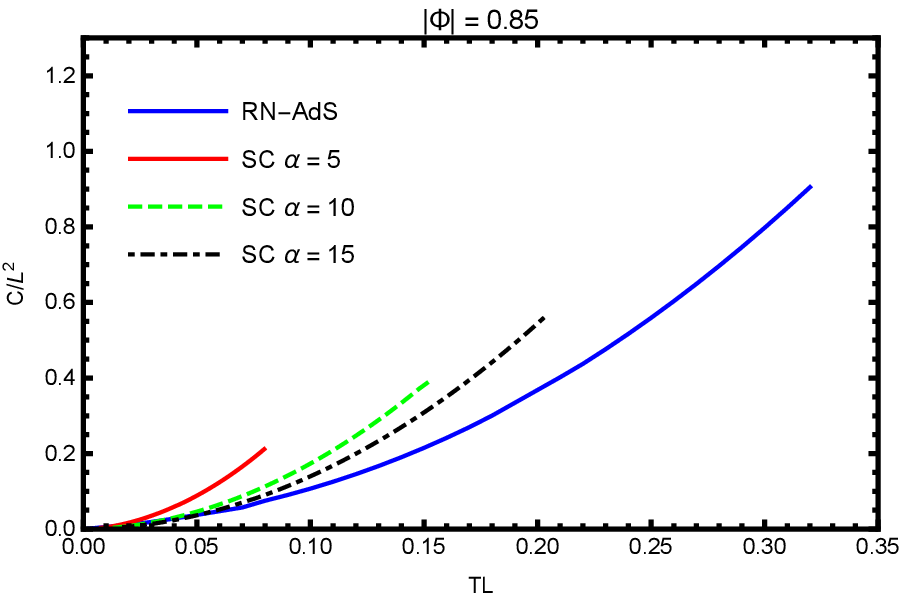}
    \includegraphics[width = 8. cm]{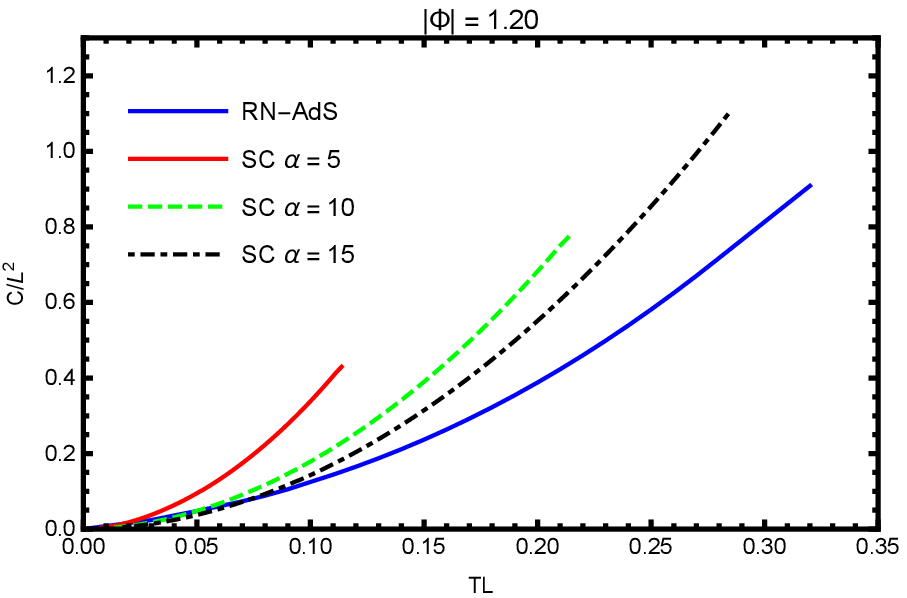}
    \caption{Show the heat capacity $C/L^2$ plot against the reduced temperature $TL$ with $L=1$.  Left: for fixed $|\Phi|=0.5$ Right: for fixed $|\Phi|=1.2$. The solid blue and red lines are RN-AdS and scalarized solution with $\alpha=5$. The dashed green and dot-dashed black lines are scalarized solution with $\alpha=10$ and $15$ respectively. }
    \label{fig: Heat capacity vs temp}
\end{figure}

 In Fig.~\ref{fig: Heat capacity vs temp}, we plot the reduced heat capacity $C/L^2$ as a function of the reduced temperature $TL$ with corresponding to the temperature profiles in Fig.~\ref{fig: Temp vs r+} for the SC BH branch (red, dashed green and dot-dashed black curves) compared with the RN-AdS BH (blue curve). 
 The heat capacity of these two branches are both positive and increased when the increasing temperature, so these two branches are locally in thermal equilibrium against microscopic fluctuations. 
 The behavior of the heat capacity of these two branches are significantly different, namely, the heat capacity of the SC BH is larger than the RN-AdS BH and reaches its maximum value at $T_B$ where $T>T_B$ SC BHs are no longer exist. On the other hand, the heat capacity of the RN-AdS BH increases monotonically with when $T>0$. 
 
As shown in Fig.~\ref{fig: Temp vs r+}, when $r_+<r_e$, where $r_e$ denotes the horizon radius of extremal RN-AdS BH, there exists only SC BH while the existence of the RN-AdS BH in this region would violate the cosmic censorship conjecture \cite{Penrose:1969pc}. 
Note that the SC BH approaches the extremal limit with zero temperature comparatively slower than the RN-AdS BH since its heat capacity is greater than RN-AdS BH. 
In other words, the SC BH radiates more energy to reduce one unit of the temperature.
Remarkably, in the planar horizon case, the SC BH exists at near zero temperature up to $T_B$. In contrast to spherical horizon case, the SC BH exists at the lowest (nonzero) temperature $T_\text{min}$ up to arbitrarily high temperature \cite{Guo2022Extended}. 

\begin{figure}[H]
    \centering
    \includegraphics[width = 8. cm]{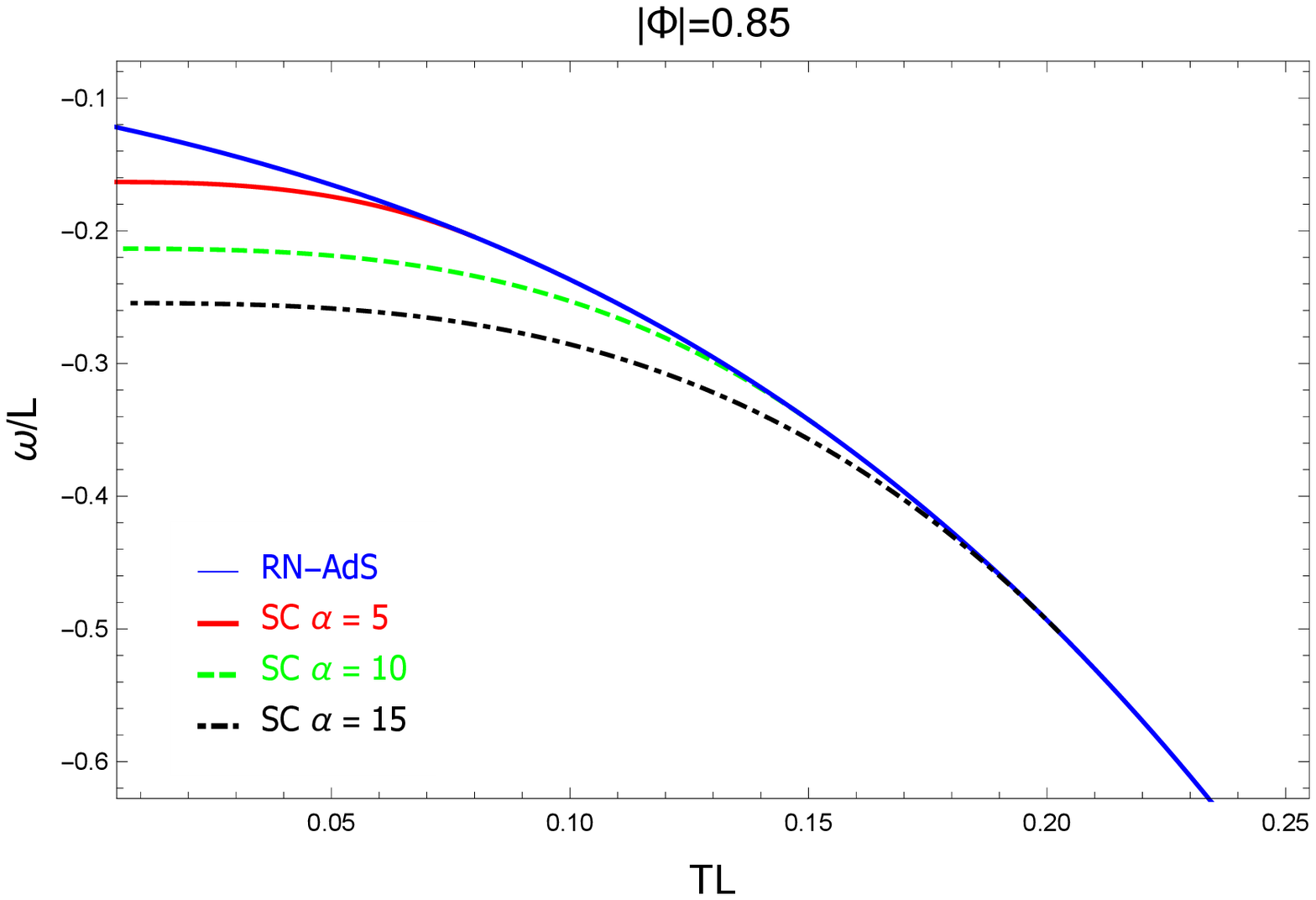}
    \includegraphics[width = 8. cm]{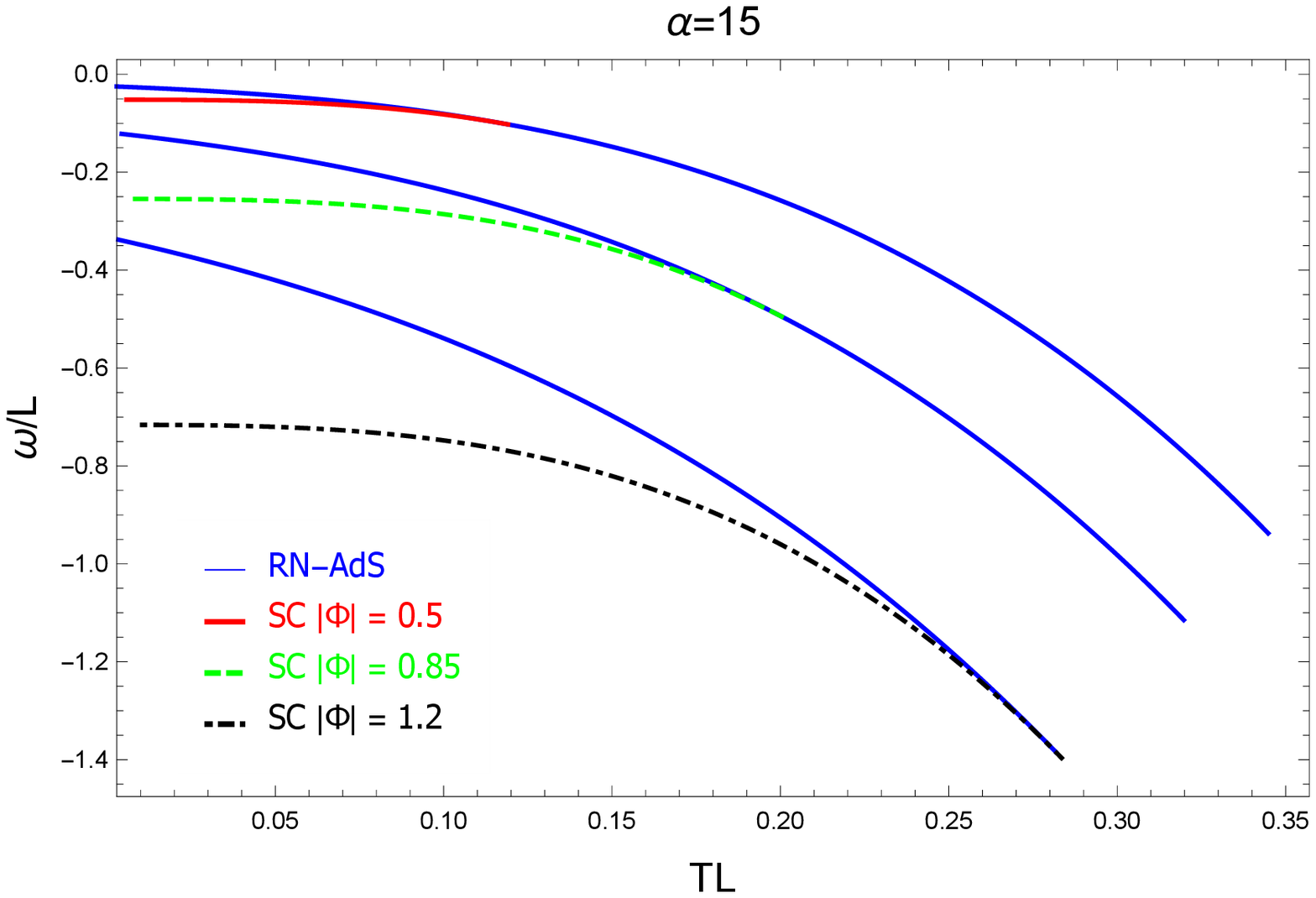}
    \caption{Grand potential density $\omega$  plots against the reduced temperature $TL$ with $L=1$. Left: for fixed $|\Phi|=0.85$ and varying $\alpha=5,10,15$ Right: for fixed $\alpha=15$ and varying $|\Phi|=0.5,0.85,1.2$.}
    \label{fig: grand vs temp}
\end{figure}

In order to investigate the global stability of BH in EMS system with fixed $|\Phi|$ ensemble, one needs to consider the grand potential. For planar black hole, the extensive quantities such as mass, entropy, charge and free energy are diverge since $\mathcal{V}_2$ has an infinite extension. Therefore it is more appropriate to use the density of these quantities as $x_i=4\pi L^2X_i/\mathcal{V}_2$ where $X_i$ represents the extensive quantities. In this way, the grand potential in Eq. \eqref{Gibbs} can be written in the form
\begin{eqnarray}
\Omega = \frac{\mathcal{V}_2}{4\pi L^2}\omega, \label{Grand potential}
\end{eqnarray}
where
\begin{eqnarray}
\omega =M-T_\text{H}s_\text{BH}-Q\Phi ,
\end{eqnarray}
defined as the grand potential density. Here $M$, $Q$ and $s_\text{BH}=\pi r^2_+$ are the mass, charge and entropy density, respectively. In Fig.~\ref{fig: grand vs temp}, we present the reduced grand potential density $\omega /L$ as the function of reduced temperature $TL$.
These graphs show that there are two branches of black hole emerged at zero temperature with different values of the grand potential: RN-AdS BH and SC BH. For the solid blue, solid red, dashed green and dot-dashed black curves correspond to RN-AdS BH and SC BH with different coupling constant $\alpha = 5, 10$ and $15$, respectively. At low temperature $T<T_B$, the thermodynamically preferred state is the SC BH since it has lower grand potential than the RN-AdS BH. 

At $T=T_B$, the SC BH and RN-AdS BH coalesce, while RN-AdS BH persists for all high temperature. Since the heat capacity, which associated to the second derivative of the free energy, i.e., $\displaystyle C=-T\left(\frac{\partial^2 \Omega}{\partial T^2}\right)_\Phi$, is discontinuous. Thus the SC - RN-AdS BH phase transition is the second order type of phase transition. In the left panel in Fig.~\ref{fig: grand vs temp}, we explore the influence of $\alpha$ on the phase structures of BH in EMS system by varying $\alpha =5, 10, 15$ for fixed $|\Phi|=0.85$. We find that the grand potential of SC BH are more negative value when the $\alpha$ increases, so the SC BH with larger $\alpha$ are more thermodynamically favored than the BH with smaller $\alpha$.
In the right panel in Fig.~\ref{fig: grand vs temp}, we fixed $\alpha =15$ and vary $|\Phi|=0.5, 0.85$ and $1.2$ respectively. The grand potential curves of SC BH and RN-AdS BH are both decreased from top to bottom when $|\Phi|$ is increasing.

\begin{figure}[H]
    \centering
    \includegraphics[width = 12 cm]{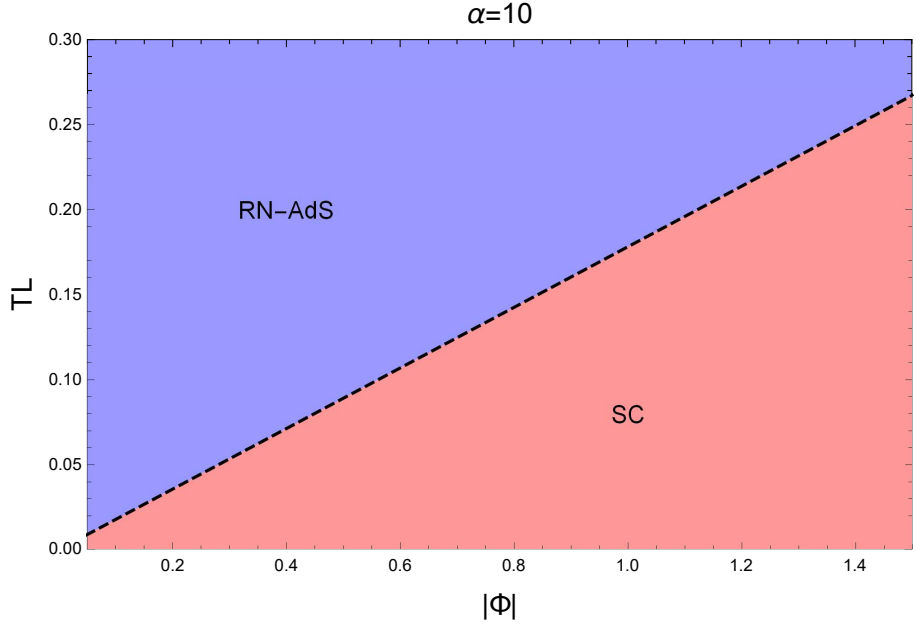}
    \caption{Temperature of black hole solutions plot against electrostatic potential for $\alpha=10$. The dashed black line indicates the temperature at the bifurcation points. The lower region is dominated by scalarized solution while the upper region belongs to RN-AdS.}
    \label{fig:TB vs Phi}
\end{figure}

Figure~\ref{fig:TB vs Phi} displays the phase diagram of the grand canonical ensemble of BH in EMS system with the lowest grand potential at fixed $\alpha =10$ in the $TL-|\Phi|$ plane. 
The dashed black curve represents the line of coexistence of the SC BH and the RN-AdS BH, where the transition across this line is of the second order type.  This occurs when the grand potential of these configurations are degenerated at $T=T_B$. In the upper region, there is only RN-AdS BH; however, when the temperature decreases to $T_B$, the RN-AdS BH develops a scalar hair near the event horizon to form the SC BH, which represents in the lower region of the coexistence line.
In this region, the SC BH can be the globally preferred thermal state.

\subsection{Fixed charge}

\begin{figure}[H]
    \centering
    \includegraphics[width = 5.4 cm]{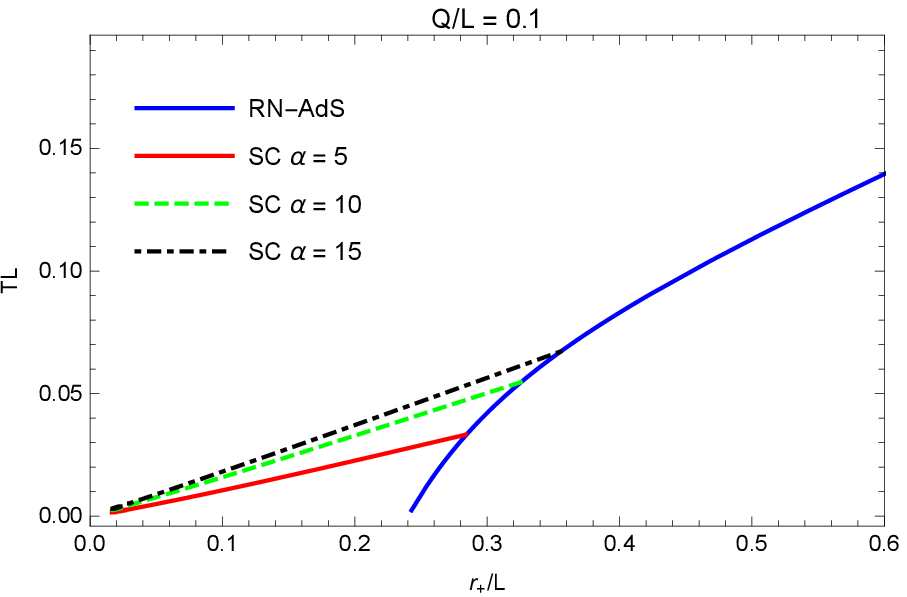}
    \includegraphics[width = 5.4 cm]{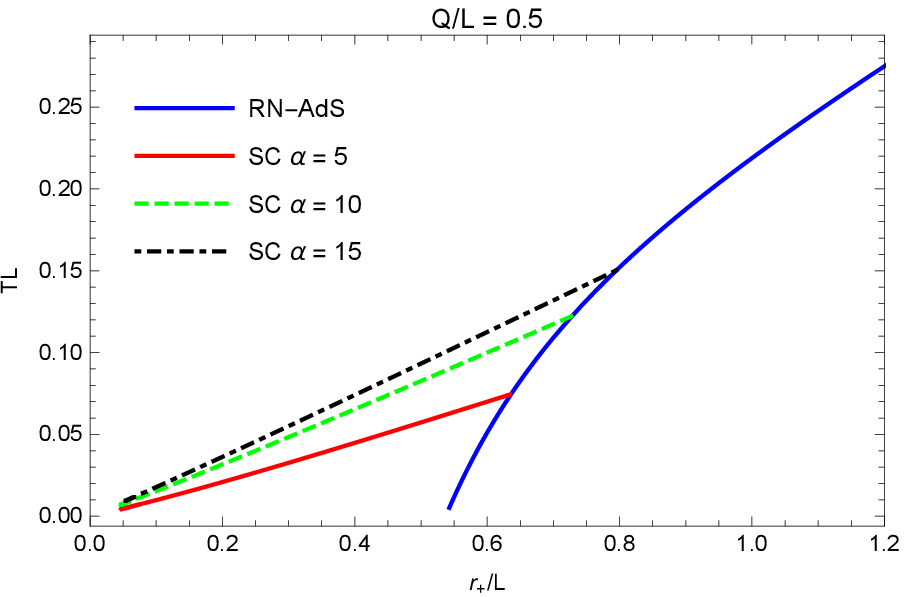}
    \includegraphics[width = 5.4 cm]{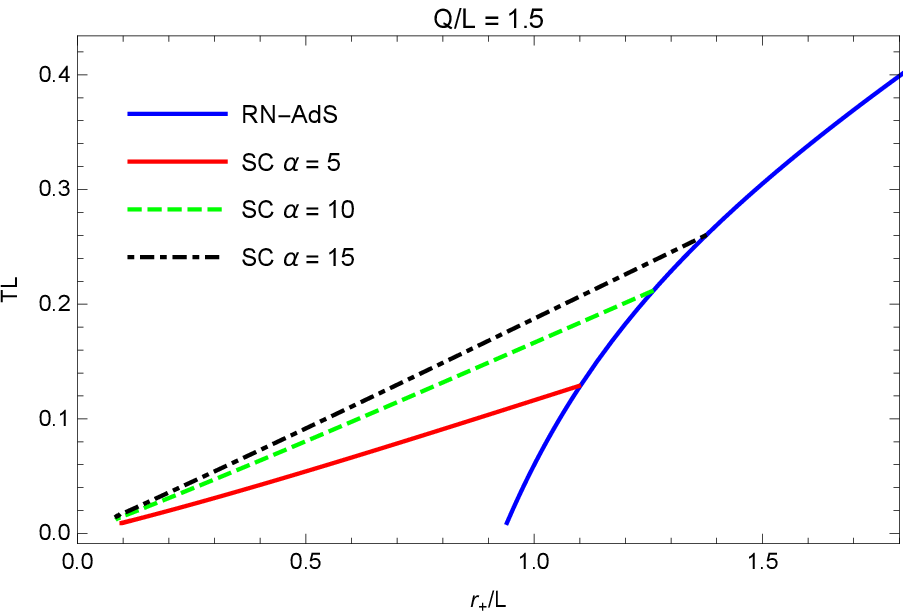}
    \caption{Plots of the reduced Hawking temperature $TL$ against the reduced horizon radius $r_+/L$ of two BH solutions in the EMS gravity with three different values of the reduced charge $Q/L = 0.1, 0.5, \text{ and } 1.5$ in the left, middle and right panels, respectively. The RN-AdS BH is identified by the solid blue curve, while the SC BHs with different $\alpha = 5, 10 \ \text{and} \ 15$ are represent in solid red, dashed green and dot-dashed black curves, respectively.}
    \label{fig:T vs r+ canonical}
\end{figure}

To study the BH thermodynamics of the EMS gravity in the canonical ensemble, we fix the bulk electric charge $Q$ of the BH solutions instead of fixing the electric potential $\Phi$. 
Fig~\ref{fig:T vs r+ canonical} shows the behavior of $TL$ as a function of $r_+/L$ with three different values of $Q/L$. 
For low temperature $T<T_B$ region, there are two branches of BH solutions of different horizon radii. 
We find that these two BH solutions with small and large horizon radii correspond to the SC BH and RN-AdS BH respectively.

\begin{figure}[H]
    \centering
    \includegraphics[width = 5.4 cm]{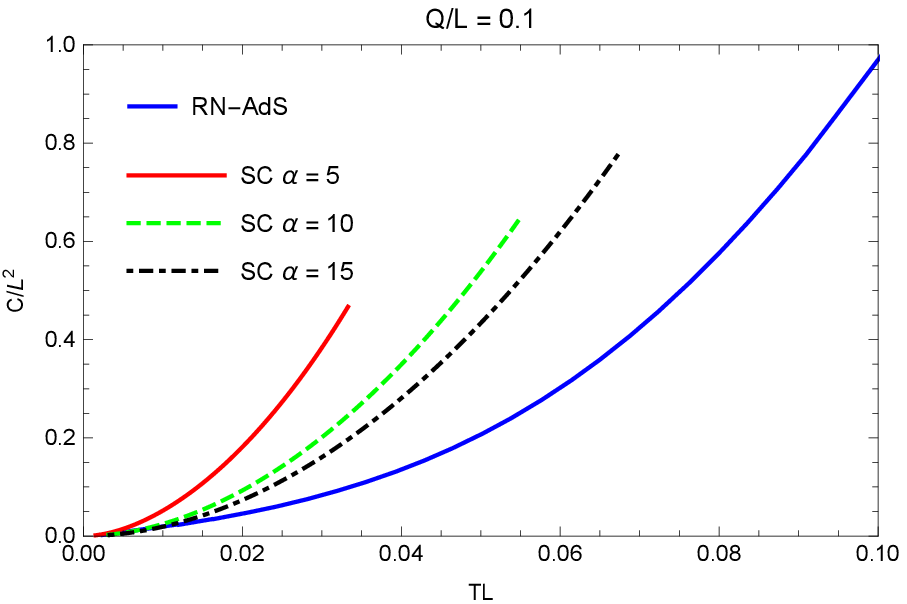}
    \includegraphics[width = 5.4 cm]{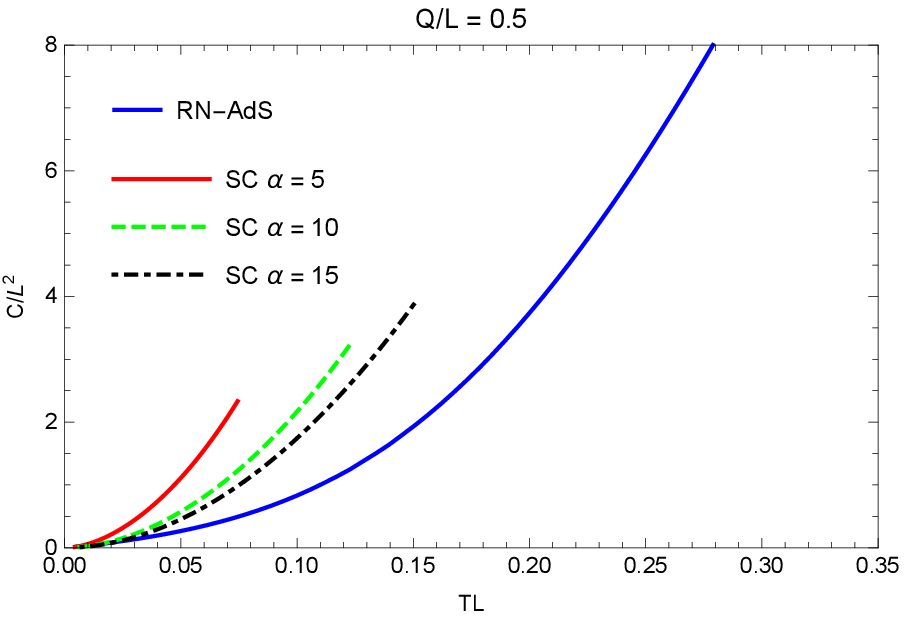}
    \includegraphics[width = 5.4 cm]{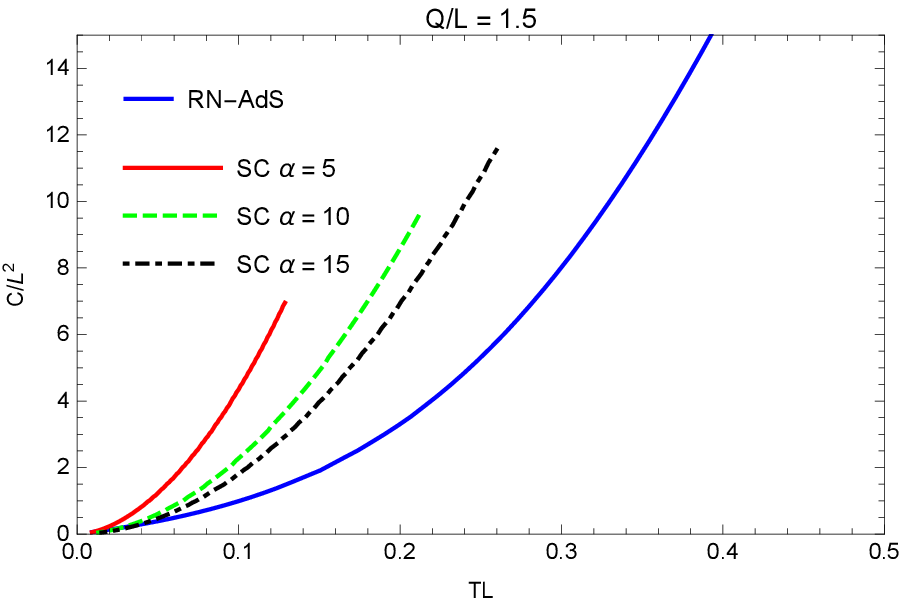}
    \caption{Plot of the reduced heat capacity $C/L^2$ as the function of the reduced temperature $TL$ for planar SC and RN-AdS BHs with the variation of $\alpha = 5, 10, \text{ and } 15$ and $Q/L = 0.1, 0.5, \text{ and } 1.5$.}
    \label{fig:HeatCapa_Temp}
\end{figure}

\begin{figure}[H]
    \centering
    \includegraphics[width = 5.4 cm]{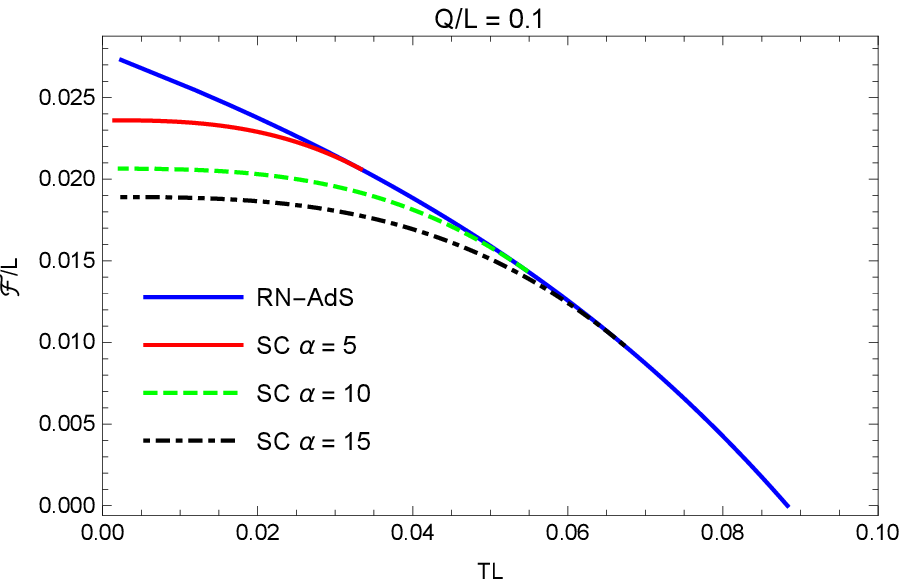}
    \includegraphics[width = 5.4 cm]{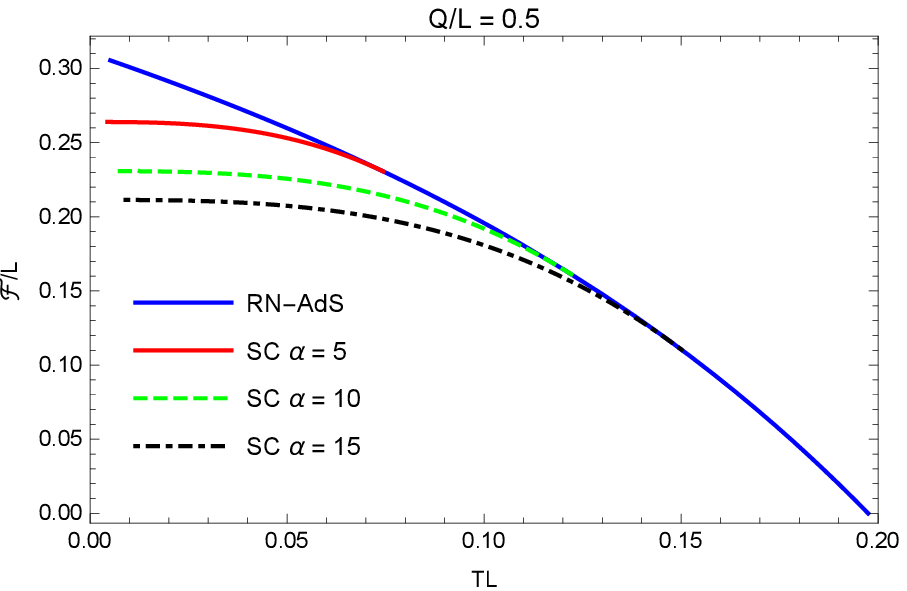}
    \includegraphics[width = 5.4 cm]{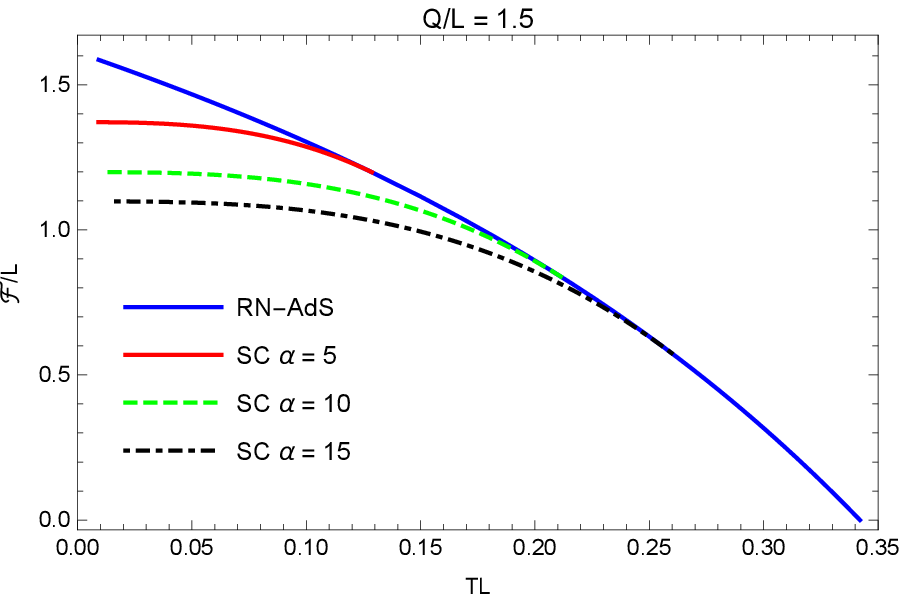}
    \caption{The Helmholtz free energy against the temperature of the phase transition between the scalarized planar black hole and the RN-AdS planar black hole with the variation of $\alpha = 5, 10, \text{ and } 15$ and $Q/L = 0.1, 0.5, \text{ and } 1.5$.}
    \label{fig:FreeHelm_Temp}
\end{figure}

In Figs~\ref{fig:T vs r+ canonical},~\ref{fig:HeatCapa_Temp},~and~\ref{fig:FreeHelm_Temp}, the RN-AdS BH is identified by the solid blue curve, while the SC BHs with different $\alpha = 5, 10 \ \text{and} \ 15$ are represented in solid red, dashed green and dot-dashed black curves, respectively. 
When the temperature increases to $T_B$, these two BHs are degenerate at the same horizon radius and SC BH transits to RN-AdS BH. Consequently, the RN-AdS BH is the only phase that exists at any high temperature regime.
By varying $Q/L = 0.1, 0.5$ and $1.5$ from the left to the right panels in Fig~\ref{fig:T vs r+ canonical}, we find that $T_B$ and its corresponding horizon radius increase when $Q/L$ increases.
At each value of $Q/L$, we also compare the effect of the strength of the coupling function between the massless scalar field and the Maxwell gauge field on the thermal behavior of SC BHs by varying $\alpha$.
These graphs indicate that $T_B$ and its corresponding horizon radius increase with $\alpha$.

To investigate the local stability of two BH solutions of the EMS gravity in the canonical ensemble, we plot the reduced heat capacity $C/L^2$ against the reduced temperature $TL$ in Fig~\ref{fig:HeatCapa_Temp} for different reduced charge $Q/L$. 
These graphs illustrate that the heat capacity of these two BH backgrounds are both positive, so they can be in thermal equilibrium with heat reservoir at fixed temperature.
However, SC BH only exists in the range $0\leq T \leq T_B$ with higher heat capacity than the RN-AdS BH and consequently disappear for $T>T_B$ region.
On the other hand, RN-AdS BH always exists at any temperature. 
For fix $\alpha$ and vary BH's charge, $T_B$ is shifted to the right when the bulk charge of the black hole increases.
Moreover, when we fix $Q$ of the black hole and vary $\alpha$, the results indicate that $T_B$ move to the right when $\alpha$ is larger.

A free energy corresponding to the canonical ensemble is the Helmholtz free energy, which is denoted by $F$. 
For planar BH, it diverges in the same way as the grand potential of the fix $\Phi$ ensemble. Thus, we use the Helmholtz free energy density $\mathcal{F}$ instead of $F$ to investigate the global stability of BHs in the EMS gravity. 
The Helmholtz free energy density is defined as
\begin{eqnarray}
F = \frac{\mathcal{V}_2}{4\pi L^2}\mathcal{F},
\end{eqnarray}
where
\begin{eqnarray}
\mathcal{F} = M - T_\text{H}s_\text{BH}.
\end{eqnarray}
Figure~\ref{fig:FreeHelm_Temp} displays the Helmholtz free energy as a function of temperature. 
In high temperature regime $T>T_B$, we find the planar RN-Ads BH is the most globally preferred phase in the AdS space. 
As one lower temperature to $T_B$ and keep $Q/L$ to be constant, the RN-AdS BH becomes unstable and develops an atmosphere in the form of massless scalar with finite peak near the event horizon to form the SC BH.
The transition between these two BH configurations can be classified to the second order phase transition since $C/L^2$ shows a discontinuity and $\mathcal{F}/L$ is continuous and smooth function at $T=T_B$ as shown in Fig~\ref{fig:HeatCapa_Temp} and Fig~\ref{fig:FreeHelm_Temp}, respectively.   
For low temperature $T<T_B$, the SC BH can be the globally preferred state.  
As we increase $Q/L$, both BH configurations move to lower $\mathcal{F}/L$, while $T_B$ of the system increases.
By varying $\alpha =5, 10$ and $15$ and fixed $Q/L =0.1, 0.5$ and $1.5$, respectively. we find  $\mathcal{F}/L$ of SC BHs are lower from the top to bottom.

\begin{figure}[h]
    \centering
    \includegraphics[width = 12 cm]{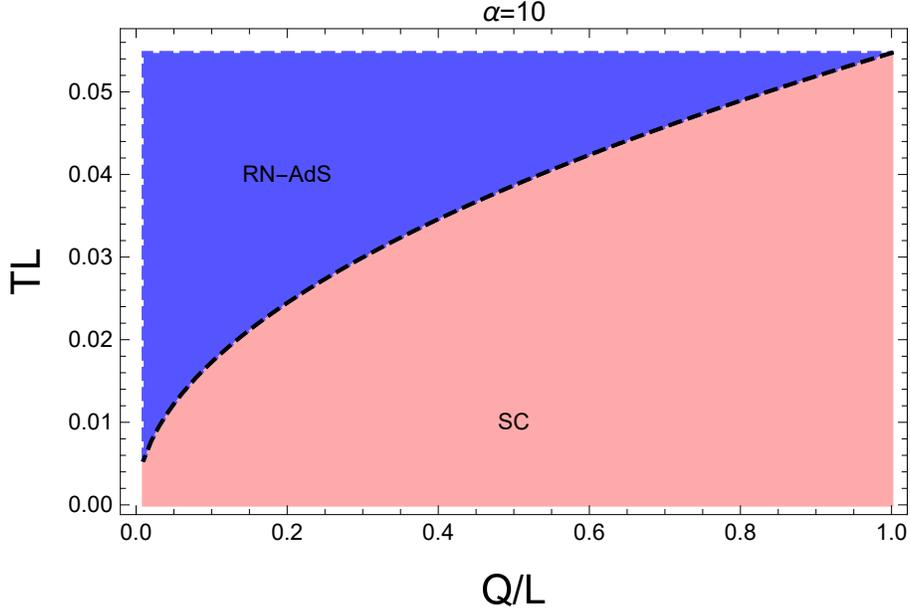}
     \caption{The plot of the temperature against the charge where the phase transition of the scalarized planar black hole emerging from the RN planar black hole with the parameter set as follows: $\alpha = 10$, $Q / L = 0.1$, and $Q \in [0.0, 1.0]$.}
    \label{fig:temp_vs_Q_canonical}
\end{figure}

For the canonical ensemble, Fig~\ref{fig:temp_vs_Q_canonical} demonstrates the phase diagram of RN-AdS and SC BHs with the lowest $\mathcal{F}/L$ in the $TL - Q/L$ plane when $\alpha=10$ . The dashed black line represents the second order phase transition dividing the RN-AdS BH (upper region) from the SC BH (lower region).

\section{Conclusions}

In this work, the spontaneous scalarization of asymptotically AdS charged black holes of planar event horizon in the  Einstein-Maxwell-calar theory with nonminimally coupling between scalar field and Maxwell field is investigated. It is tachyonic instability that drives scalar-free charged black holes away from their stability. When the condition $\mu^2_{eff} < \mu^2_{BF}$ holds, scalarized planar charged black holes emerge when charge $q$ and the coupling constant $\alpha$ are sufficiently large as illustrated in Fig.~\ref{fig:solspace}. Moreover, in the region where RN-AdS and SC BHs co-exist, we find that SC BHs are entropically preferred over RN-AdS. 
From our investigation, we observe that the SC BHs with low $\alpha$ exhibit a possibility of instability while perturbation equations of SC solutions with higher $\alpha$ do not develop any bound states. By computing quasinormal frequencies, we find that all scalarized black holes investigated in this work are linearly stable against scalar perturbations. In fact, the scalarized solutions become more stable as $\alpha$ is increased. In contrast, we show that the RN-AdS black holes do suffer from tachyonic instability. Despite that, comparing to the spherically symmetric case, topological scalarized solutions are more difficult to obtain. However, we find that their domains of existence, the reduced event horizon areas of black hole and stability behaviors share several similarities.

In terms of thermodynamic behaviors of planar BH solutions in the EMS gravity, we find two stable branches of BHs: RN-AdS BH and SC BH in both grand canonical and canonical ensembles.
There is only planar RN-AdS BH branch at high temperature regime $T>T_B$.
By decreasing temperature, planar RN-AdS BH becomes unstable below $T_B$ and develops nonzero finite value of real scalar field near its event horizon, i.e., $\phi_0\neq 0$.
The phase transition between RN-AdS BH to SC BH is the second order phase transition since the behavior of heat capacity shows a discontinuity at $T=T_B$.
Remarkably, SC BH are thermodynamically favored than RN-AdS BH in $T<T_B$ region because they have smaller free energy than RN-AdS BH.
The phase diagram in Fig~\ref{fig:TB vs Phi} and Fig~\ref{fig:temp_vs_Q_canonical}  display a region where RN-AdS BH and SC BH exist with the lowest free energy in grand canonical and canonical ensemble, respectively.
Unlike the spherical SC BHs that have no extremal limit since they only exist from nonzero temperature up to arbitrary temperature, the planar SC BHs emerge from the bifurcation point at $T_B$ which is thermodynamically preferred than RN-AdS BH. Below $T_B$, planar SC BH approaches toward the extremal limit. The spherical BHs in the EMS model has richer thermodynamic phase structure and phase transition than the planar horizon case.
In the spherical case, the BH configuration exhibits a phase transition in a similar way as the liquid-gas one at small $Q$ regime, while the reentrant phase transition can occurs in some range of parameters in the large $Q$ limit.
Nevertheless, the thermal second order phase transition between planar RN-AdS BH and SC BH in the EMS model is reminiscent of normal conductor-superconductor phase transition.  
This similarity may demonstrate as follows.
In the gauge/gravity duality, one can interpret the radial coordinate $r$ of AdS space as geometrical view of the renormalisation group flow.
In other words, running the coordinate $r$ from the boundary to the interior of the bulk in AdS corresponds to flowing down the energy scale from high energy (UV) to low energy (IR) of the field theory.
In the left panel of Fig.~\ref{fig:sols} shows a scalar field profile in the bulk spacetime. 
The scalar field is concentrated in the deep interior of spacetime and monotonically decreases toward zero amplitude near the boundary.
In thermodynamics perspective, SC BH with $\phi_0\neq 0$ will develop in the low temperature limit $T<T_B$, however, for $T>T_B$ the scalar field's amplitude vanishes $\phi_0=0$, at this point the SC BH transits to scalar-free BH.   
Such mechanism implies a spontaneous symmetry breaking of $U(1)$ gauge symmetry in the IR limit and the symmetry is restored at the UV physics.
This suggests that SC BH and RN-AdS BH in the EMS model might behave as a superconducting phase and normal conducting phase, respectively, as in the Abelian-Higgs model of superconductivity \cite{Weinberg:1986cq}.


As a general extension of this work, it is interesting to explore the dependence of these scalarized black holes on various form of coupling functions where the similar work is done with spherical horizon \cite{Fernandes:2019rez}. A nonlinear dynamical evolution of scalarized spherical black hole is studied in \cite{Xiong:2022ozw}. Similar should be done for the planar case where it should offer more insight into perturbative stability of scalarized planar black holes. Given, thermodyanmics behaviors and phase structure of our scalarized solutions, it would be extremely interesting to explore these solutions in the context of holographic superconductor. We leave these for future works.

\begin{acknowledgments}

The authors would like to thank Carlos A. R. Herdeiro, Alexandre M. Pombo and Napat Poovuttikul for very useful discussions. This work (Grant No. RGNS 64-217) was supported by Office of the Permanent Secretary, Ministry of Higher Education, Science, Research and Innovation  (OPS MHESI), Thailand Science Research and Innovation (TSRI) and Silpakorn university. T. Tangphati was supported by King Mongkut's University of Technology Thonburi's Post-doctoral Fellowship. E. Hirunsirisawat acknowledges the financial support provided by the Center of Excellence in Theoretical and Computational Science (TaCS-CoE), KMUTT. Moreover, this research project is supported by Thailand Science Research and Innovation (TSRI) Basic Research Fund: Fiscal year 2023 under project number FRB660073/0164.


\end{acknowledgments}

\appendix

\section{MASS AND CHARGE OF PLANAR EMS SOLUTION}\label{Komar integral}

In this appendix, we shall derive mass and electric charged of planar BH solutions in the EMS gravity via the Komar integral, which we have used them to derived some thermodynamic quantities of BHs. 
Let's first consider, the Maxwell's equation with an external current $J^\mu$. Thus Eq.~\eqref{MW} becomes
\begin{eqnarray}
\nabla_{\nu}\left(\mathcal{G}F^{\mu\nu}\right) = J^\mu,
\end{eqnarray}
where $\mathcal{G}=e^{\alpha \phi^2}$. The electric charge $\Tilde{Q}$ passing through a spacelike hypersurface $\Sigma$ can be defined by an integral over spatial coordinates $x^i$ on $\Sigma$
\begin{eqnarray}
\Tilde{Q} = -\frac{1}{4\pi}\int_\Sigma d^3x\sqrt{\gamma^{(3)}}n_\mu J^\mu =- \frac{1}{4\pi}\int_\Sigma d^3x\sqrt{\gamma^{(3)}}n_\mu \nabla_\nu \left(\mathcal{G}F^{\mu\nu}\right),
\end{eqnarray}
where $\gamma^{(3)}_{ij}$ is the induced metric and $n_\mu$ is the unit normal vector associated to $\Sigma$. 
By using the Stokes's theorem, we can rewrite the volume integral in the above relation as a surface integral
\begin{eqnarray}
\Tilde{Q}=-\frac{1}{4\pi}\int_{\partial \Sigma}d^2x\sqrt{\gamma^{(2)}}n_\mu \sigma_\nu \mathcal{G}F^{\mu \nu}, \label{Physical charge}
\end{eqnarray}
where $\gamma^{(2)}_{ij}$ is the induced metric and $\sigma_\mu$ is the unit normal vector associated to $\partial \Sigma$. 
For planar EMS solutions in Eq.~\eqref{metric}, the surfaces $\Sigma$ and $\partial \Sigma$ are the constant-time hypersurface and $\mathbb{R}^2$ surface at spatial infinity, respectively. Therefore the unit normal vectors associated to $\Sigma$ and $\partial \Sigma$ are
\begin{eqnarray}
n_\mu = \left( -N^{1/2}(r)e^{-\delta (r)}, 0, 0, 0\right), \ \ \ \text{and} \ \ \ \sigma_\mu = \left( 0, N^{-1/2}(r), 0, 0 \right), \label{normal vector}
\end{eqnarray}
respectively. 
The line element on surface $\partial \Sigma$ is
\begin{eqnarray}
\gamma^{(2)}_{ij}dx^idx^j = \frac{r^2}{L^2}\left( dx^2 + dy^2 \right),
\end{eqnarray}
and the volume element takes the form
\begin{eqnarray}
d^2x\sqrt{\gamma^{(2)}} = \frac{r^2}{L^2}dxdy. \label{Volume element}
\end{eqnarray}
From definition of normal vectors in Eq.~\eqref{normal vector}, one can compute
\begin{eqnarray}
n_\mu \sigma_\nu \mathcal{G}F^{\mu \nu} = e^{-\delta (r)}\mathcal{G}V'(r). \label{Integrand}
\end{eqnarray}
Substituting Eq.~\eqref{Volume element} and Eq.~\eqref{Integrand} into Eq.~\eqref{Physical charge} gives
\begin{eqnarray}
\Tilde{Q} = -\frac{\mathcal{V}_2}{4\pi L^2}\left( r^2e^{-\delta (r)}\mathcal{G}V'(r) \right)_{r\rightarrow \infty},
\end{eqnarray}
where $\mathcal{V}_2$ denotes spatial area of constant $r$ surface.
Remark that the electric charge $\Tilde{Q}$ is evaluated at spatial infinity. 
By using Eqs.~\eqref{Bcinf2}-\eqref{Bcinf4}, we obtain
\begin{eqnarray}
\Tilde{Q} = \frac{\mathcal{V}_2 Q}{4\pi L^2}.
\end{eqnarray}

In an asymptotically AdS space, the physical mass can be measured with respect to a reference AdS background. In this way, the Komar mass can be written in the following form
\begin{eqnarray}
E = \frac{1}{4\pi}\int_{\partial \Sigma}d^2x\sqrt{\gamma^{(2)}}n_\mu \sigma_\nu \nabla^\mu K^\nu - E_\text{AdS}. \label{Komar mass}
\end{eqnarray}
We define the time-translation Killing vector $K^{\nu} \equiv (1,0,0,0)$,
$n_\mu$ and $\sigma_\mu$ are defined in Eq.~\eqref{normal vector}.
The Komar integral of vacuum AdS space $(E_\text{AdS})$ is given by
\begin{eqnarray}
E_\text{AdS} = \frac{\mathcal{V}_2r^3}{4\pi L^4}.
\end{eqnarray}
Therefore, we have
\begin{eqnarray}
n_\mu \sigma_\nu \nabla^\mu K^\nu = \frac{1}{2}e^{-\delta(r)}N'(r)\left( 1-2\delta'(r) \right).
\end{eqnarray}
Thus the first term of \eqref{Komar mass} becomes
\begin{eqnarray}
\frac{1}{4\pi}\int_{\partial \Sigma}d^2x\sqrt{\gamma^{(2)}}n_\mu \sigma_\nu \nabla^\mu K^\nu &=& \frac{\mathcal{V}_2}{4\pi}\left[ \frac{r^2}{2L^2}e^{-\delta (r)}N'(r)\left( 1-2\delta'(r) \right) \right], \nonumber \\
&=& \frac{\mathcal{V}_2}{4\pi}\left[ \frac{M}{L^2} - \frac{Q^2}{2L^2r} + \frac{r^3}{L^4} + \mathcal{O}(r^{-2}) \right].
\end{eqnarray}
Evaluating at the boundary ($r\rightarrow \infty$) with $E_\text{AdS}$, we obtain the Komar mass of planar BH solutions as
\begin{eqnarray}
E = \frac{\mathcal{V}_2M}{4\pi L^2}.
\end{eqnarray}

\bibliography{EMS}

\end{document}